\documentclass[preprint,12pt]{elsarticle}

\usepackage{amssymb}
\usepackage{amsmath}
\usepackage{array}
\usepackage{booktabs}
\usepackage{caption}
\usepackage{float}
\usepackage{graphicx}
\usepackage[labelfont=bf, labelsep=period]{caption}
\usepackage{makecell}
\usepackage{multirow}
\usepackage{siunitx}
\usepackage{tablefootnote}

\usepackage[left]{lineno}

\journal{J. Manuf. Process.}

\begin{document}
% \linenumbers 
\begin{frontmatter}



\title{Self-Focusing Control for Depth-Precise Wafer Slicing of 4H-SiC in Femtosecond Laser Processing}

%% author
\author[a]{Dong Hee Kang}
\author[b,c]{Jaeseung Lim}
\author[a]{Mishfaqur Rahman}
\author[b]{Seongheum Han}
\author[b]{Jae-Hak Lee}
\author[b,c]{Seungman Kim\corref{cor1}}
\author[a]{Jihoon Jeong\corref{cor1}} 

\affiliation[a]{organization={Wm Michael Barnes '64 Department of Industrial \& Systems Engineering, Texas A\&M University},%Department and Organization
            %addressline={3131 TAMU}, 
            city={College Station},
            state={TX 77843},
            country={United States}}

\affiliation[b]{organization={Semiconductor Manufacturing Research Center, Korea Institute of Machinery and Materials (KIMM)},
            %addressline={00000}, 
            city={Daejeon 34103},
            country={Republic of Korea}}
            
\affiliation[c]{organization={Department of Robot $\cdot$ Manufacturing Systems, University of Science and Technology (UST)},
            %addressline={00000}, 
            city={Daejeon 34113},
            country={Republic of Korea}}  
            
\cortext[cor1]{Corresponding authors:  jihoonjeong@tamu.edu \& kimsm@kimm.re.kr}
            
%% Abstract
\begin{abstract}
4H-SiC has emerged as a third-generation chip material because its superior thermal conductivity and high breakdown field enable the material to achieve high power density and higher switching frequencies in power-electronics applications. As chip architectures evolve toward 3D and heterogeneous integration, the mechanical and thermal design space tightens while yield risks grow. In particular, advanced packages require mid-process wafer thinning to < 100 $\mu$m to shorten interconnects and control thermo-mechanical stress. Femtosecond laser slicing for 4H-SiC wafers offers a non-contact processing approach to produce thin layers with low defects, while strong optical nonlinearities obscure the relationship between the laser parameters and the resulting slicing quality. 
Here, we systematically investigate Kerr-induced self-focusing using a femtosecond laser in 4H-SiC slicing by combining experiments, a semi-empirical analytical model, and numerical ray optics simulations. We demonstrate that the interplay between pulse energy and processing depth governs the self-focusing behavior, which directly correlates with post-separation surface texture parameters and separation stress, thereby linking nonlinear beam propagation to slicing quality. Based on this relationship, we define a processability map in the pulse energy with self-focusing depth space over a normalized irradiance background. Analytically, the model extends the Marburger formula to focused beams by replacing the power ratio with a normalized irradiance. Ray optics simulations capture the geometric features at the self-focusing point and are validated against experimental observations. Within physically defined thresholds, the processability map directly connects laser parameters to separation stress and surface texture metrics, providing practical guidance for depth control beyond trial-and-error.
\end{abstract}

% \includegraphics[width=\linewidth]{GA.pdf}

% \begin{highlights}
% \item Non-contact femtosecond slicing of 4H-SiC with controlled depth
% \item Analytical model of the modified Marburger equation extended to focused beams
% \item Ray optics simulation model to quantify the beam irradiance profile
% \item Processability map organizing with pulse energy versus self-focusing depth
% \item Data cluster direct linkage between processing conditions and slice quality
% \end{highlights}

%% Keywords
\end{frontmatter}

\section{Introduction}
\label{sec1}
As demand for next-generation chips grows, wafers with superior thermophysical properties and reliability are required to support higher power density and higher switching frequencies across the power-electronics market \cite{Sahu25}. Current chip-level heat fluxes already compel the adoption of high thermal conductivity substrates. Recent CPU and GPU characterizations describe hotspot heat fluxes around 100 W cm$^{-2}$, even before considering the non-uniform heat distribution at the core sublayer \cite{Elli22}. In addition, power dissipation requires over 1,000 W cm$^{-2}$ in the high-power radio frequency at the wafer level \cite{Bar19}. Among wide-bandgap substrates, 4H-SiC is an attractive semiconductor material due to its high breakdown field and high thermal conductivity, enabling efficient heat extraction and voltage handling at high junction temperatures $\geq$ 200 $^{\circ}$C \cite{Bar19, Liu19}. 

As chip architectures evolve from planar chips to 3D and heterogeneous systems as high-bandwidth memory (HBM) stacks, the mechanical and thermal design margins become increasingly constrained, while the associated yield risks continue to rise \cite{Mo23}. These semiconductor packages rely on thin wafers $\leq$ 50 to 100 $\mu$m to shorten interconnects with controlling thermo-mechanical stress, which requires mid-processing for wafer thinning to maintain throughput and yield across lithography and backside processes \cite{Domk17}. The conventional slicing method does not achieve constraints alone. The diamond wire sawing introduces inevitable kerf loss and subsurface damage with brittle chipping, which has a trade-off between throughput and damage remains fundamental \cite{Chen21, Qiu22}. Here, non-contact pulsed-laser wafer slicing has been investigated as an alternative approach, providing a platform to study localized energy deposition and nonlinear optical effects while mitigating direct mechanical contact \cite{Domk17}.

Wafer slicing with an ultrashort pulse laser has been explored that internal modification layers can be generated in a 4H-SiC wafer with reduced thermal load. The first experimental attempt at 4H-SiC wafer slicing used a laser with a pulse duration of pico- to femto-seconds for the double pulses \cite{Kim17}. The modification layers can be formed and separated at low stress. The time delay between pulses of under 10 ps suppressed axial filament elongation and stabilized a thin modified layer suitable for separation. In a single-pulse laser slicing, the multi-focusing effect along a beam axis damages layer planarity. The ultrashort pulse laser evokes the dynamic interplay of Kerr self-focusing with plasma-induced defocusing and nonlinear absorption \cite{Coua07}. While such filamentation has been influenced for stealth dicing, where modification in depthwise is advantageous. Effective dicing with pulse laser benefits from a non-thermal beam pathway that prefers short pulses and minimal heat diffusion \cite{Xie24, Wang22-1, Wang22-2}. Following a rise in peak irradiance, strengthening optical nonlinearity can degrade layer planarity and narrow the process window \cite{Zhan23}. To improve the processing efficiency, a coupling of the simulation and experiment has been built to visualize the electric field distribution description of the nonlinear effect with the lattice temperature within the medium \cite{Wang22-1, Song24}. 

Recent research in laser processing for wafer slicing focused on the feasibility of obtaining separable layers while limiting kerf loss and surface damage across device and wafer scales, including picosecond laser-induced micro-explosion slicing \cite{Han22}, and slicing the 6-inch wafers considering crystal orientation \cite{Yao25}. For the processability, pulse laser slicing was repeated or combined with a continuous wave (CW) laser process. Process-window extension via low-energy repetition strategies to restrict micro-crack growth and reduce stress and roughness \cite{Xian24}. In the hybrid laser processing, Seed cracks were generated with low-energy picosecond laser pulses, then completed separation by CW laser-assisted splitting \cite{Jian24}. To enhance surface quality after wafer slicing, an additional laser process on the modified layer was performed for the polishing \cite{Li24}. Electrochemical exfoliation after slicing could separate with low stress at the micro-crack region at the modified layer, etching for the sub $\mu$m roughness surface quality \cite{Geng23}. However, predictable wafer slicing with a femtosecond pulse laser is still required to quantify the nonlinear collapse and its mapping to slicing to enable precise thickness control and robust processability without additional steps.

In this research, we control the slicing depth while keeping the quality of the laser-modified layer in 4H-SiC. Processability emerges on the normalized irradiance colormap as the joint trend of pulse energy and self-focusing depth, which recasts the beam power and the geometric focus interacting with properties of a medium into an applicable operating window. 
The processing conditions are intentionally fixed, not to claim global optimality across all laser slicing regimes, but to isolate and clarify how the self-focusing depth governs processability within a representative non-thermal operating window. 
The self-focusing effect is explained by experiments with an analytical model and ray optics simulations to quantify the beam irradiance profile, the self-focusing depth, and the above-threshold depth span. The analytical model extends the Marburger formula to the focused beams and defines irradiance, yielding a trend for the self-focusing depth. At the first collapse of the self-focusing point, ray optics simulations and experiments reveal that the feature dimensions (width and depth) and axial depth exhibit energy-dependent offsets relative to the geometric focus. The physical descriptors are linked to macroscopic metrics of the separation stress and the surface texture characteristics, enabling prediction and tuning of slicing quality beyond trial-and-error. 

\section{Methods}
\label{sec2}
Characteristics of the femtosecond laser slicing in 4H-SiC were investigated with experimental results, comparing the analytical model based on a semi-empirical equation and the optics simulation considering the nonlinear effect of the pulse laser. Fig. 1a,b describes the schematic of laser slicing and the subsequent layer separation. In the laser slicing, all of the processing conditions were fixed without beam irradiance power ($P_\mathrm{in}$) and geometric focal depth ($z_\mathrm{f}$). 
Separation stress ($\sigma$) and the surface texture parameters: areal-averaged surface roughness ($S_\mathrm{a}$), areal root mean square surface roughness ($S_\mathrm{q}$), surface height skewness ($S_\mathrm{sk}$), and kurtosis ($S_\mathrm{ku}$) were used as evaluation indices for the process quality.
First, a depth-resolved beam irradiance profile is described along the depth to see the contribution of the beam diameter change to the focusing point in Fig. 1c. The high irradiance causes the nonlinear effect over a critical power for the material, and an actual processing point is generated upstream of the geometric focal point in Fig. 1d. The ray optics simulation result is visualized to describe the self-focusing effect in Fig. 1e with the refractive index including the Kerr term, and the beam path changing due to the over-threshold irradiance itself. The processability map in Fig. 1f depicts the effective processing zone (EPZ) for laser slicing, which is validated with analytical and simulation models.

\begin{figure}[h!]
\centering
\includegraphics[scale=0.35]{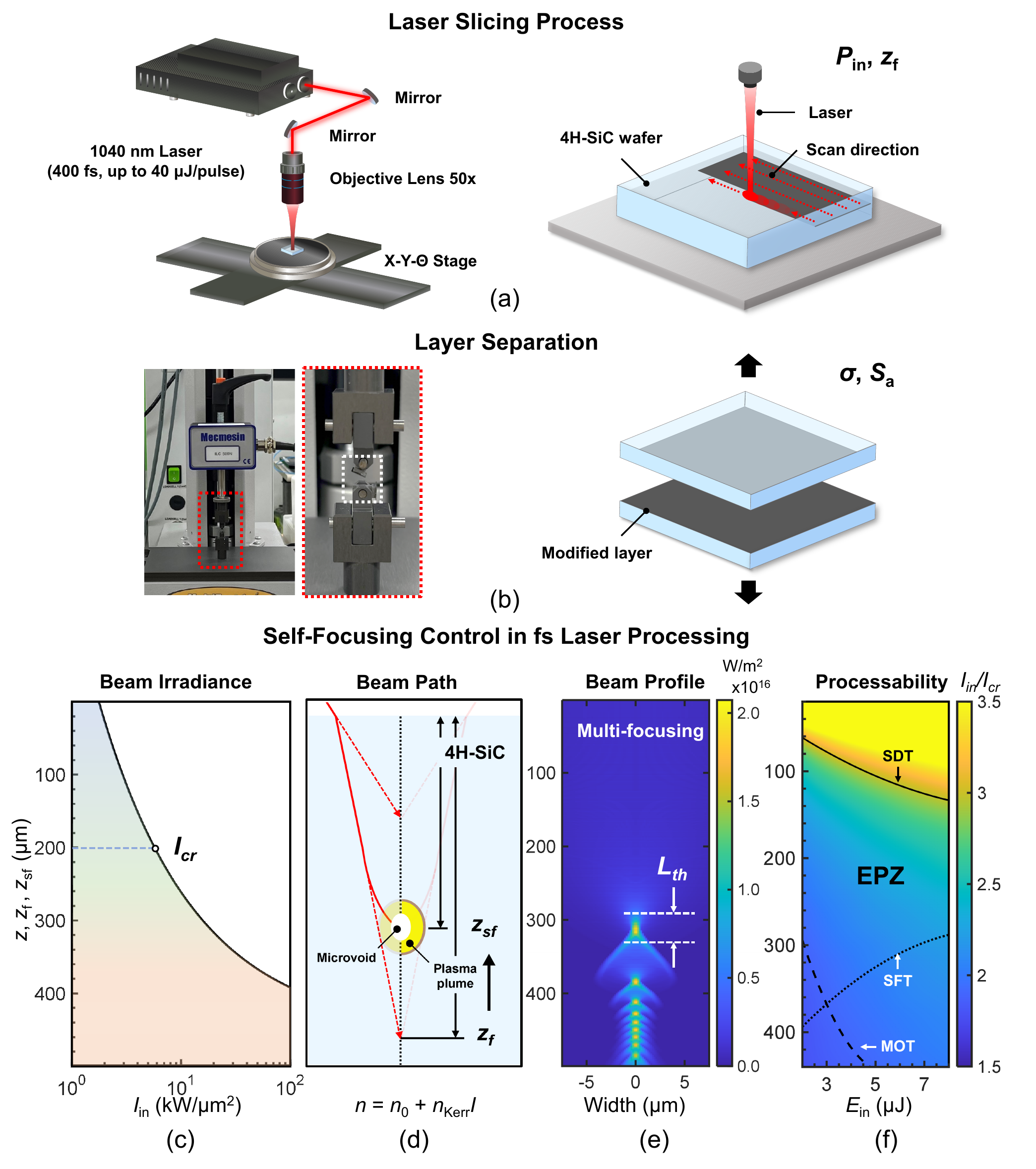}
\caption{(a) Schematics of femtosecond laser slicing in 4H-SiC and (b) the layer separation. (c) Irradiance profile by incident beam power ($P_\mathrm{in}$) along the depth ($z$) of the wafer from the analytical model. (d) Schematic of beam path with the Kerr-induced self-focusing and plasma generation at the self-focusing depth ($z_\mathrm{sf}$), upstream of the geometric focal depth ($z_\mathrm{f}$). (e) Beam profile from ray optics simulation model to evaluate experimental results with $z_\mathrm{sf}$, and above-threshold depth span ($L_\mathrm{th}$). (f) Processability map to describe the effective processing zone (EPZ) for optimal processing qualities with a low separation stress ($\sigma$) and low areal-average surface roughness ($S_\mathrm{a}$).}
\label{fig1}
\end{figure}

The wafer slicing process was performed with a focused beam path in the medium, drawing parallel single lines in a sequence to fill the target area. The crystalline structure of the 4H-SiC is decomposed to amorphous carbon and silicon at a modified layer by the focused beam plasma, which causes propagating cracks in the planar direction and separates into thin wafers. The extremely high peak power density of the femtosecond laser results in a strong nonlinear interaction, as in Appendix A. The beam experiences self-focusing upstream of the intended focal point, making it difficult to maintain precise control over the modification depth inside the material.
The interaction between femtosecond laser irradiation and the medium was elucidated through analytical and simulation models, which enable practical applications in high-precision wafer separation.

\subsection{Experimental conditions for femtosecond laser slicing of 4H-SiC wafer}
\label{2.1}
The 4-inch 4H-SiC wafer (TankeBlue Co.) was supplied as a dummy-grade substrate with a nominal thickness of $500 \pm 25 \mu$m. The manufacturer specifies a polished Si-face roughness of $R_\mathrm{a} \le 1~\mathrm{nm}$ and a CMP-finished surface roughness of $R_\mathrm{a} \le 0.2~\mathrm{nm}$, which defines the baseline condition prior to slicing. 
The wafer was prepared to 5×5 mm$^{2}$ sizes before the laser slicing. The laser slicing process was performed using a pulse laser (Lasernics, Korea) on an X-Y-$\Theta$ stage. 
The laser beam quality ($M^2$) was nearly Gaussian ($M_x^2 = 1.09$, $M_y^2 = 1.04$), which is sufficiently close to $M^2 \approx 1$ for the Gaussian-based focusing analysis.
The pulse laser with a wavelength ($\lambda_0$) of 1040 nm was used to slice the wafer, operating at a pulse duration ($\tau_l$) of 400 fs and a repetition rate ($f$) of 200 kHz. The beam diameter was 2 mm at the exit of the objective lens. The laser beam focusing depth was controlled by a z-stage with a 0.42 numerical aperture (NA) of ×50 objective lens (Mitutoyo). The laser scanning direction was decided by the crystal orientation of the 4H-SiC. Beam path lines were drawn perpendicular to the secondary flat (a-plane) of the 4H-SiC, along the [$\bar{1}\bar{1}$20] direction with a 75 mm/s scanning speed. The lines with a 30 $\mu$m scanning pitch were formed perpendicular to the primary flat (m-plane) of the 4H-SiC, along the [$\bar{1}$100] direction.  

\subsection{Laser beam propagation and focusing in a Kerr-nonlinear medium}
\label{2.2}

\begin{figure}[h!]
\centering
\includegraphics[scale=0.5]{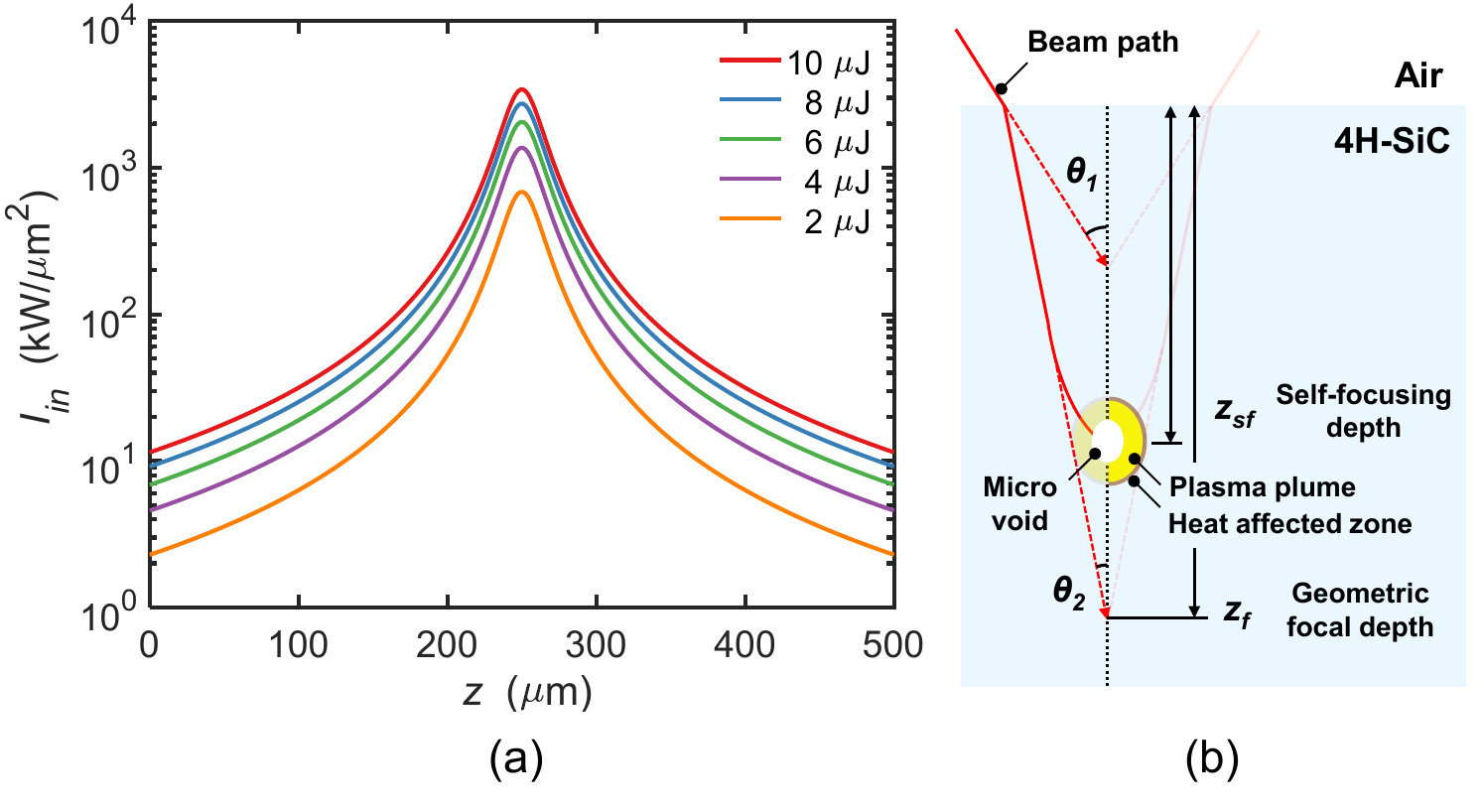}
\caption{Conceptual irradiance profile of the focused femtosecond laser beam. The profile depicts a continuous beam propagating without interaction with the medium, illustrating the beam geometry and focusing behavior, despite the actual femtosecond pulse having a very short propagation length. (a) Irradiance distributions along the propagation to depthwise for various pulse energies with the focal point positioned at the center (250 $\mu$m) of the medium (500 $\mu$m) depth. (b) Schematic illustration of the femtosecond laser beam path in a medium. The $z_f$ of the linear-optics focus inside 4H-SiC, controlled by objective $z$-translation and index-corrected to in-medium depth.}
\label{fig2}
\end{figure}

Before analyzing the laser-material interaction in 4H-SiC, we established the magnitude of the in-medium power density delivered by the focused beam. Fig. 2a describes the conceptual irradiance profile of the focused femtosecond laser beam. The profile assumed a continuous beam propagating without interaction with the medium. The irradiance of the laser beam was derived from the pulse intensity $I$ \cite{Coua07, Boyd08},
\begin{equation}
I(z)= \frac{P \cdot T}{\pi w(z)^2 \cdot f \cdot \tau_l},
\end{equation}
where $P$ is the average power of the incident laser, and $T$ is the transmittance of the laser into the 4H-SiC. The $\pi w(z)^2$ is the spot area at a specific depth ($z$) perpendicular to the beam axis of the ideal cone shape of the incident focused laser beam. The Gaussian beam radius ($w(z)$) is calculated with the beam waist radius ($w_0$) at focal depth ($z_0$), and the ratio of the specific depth to the focal depth distance in the medium ($z-z_0$) to the Rayleigh range of an ideal Gaussian beam ($z_\mathrm{R}$). The refractive index of 4H-SiC ($n_0$) is 2.55 for the 1040 nm wavelength of the incident light, which decides the Rayleigh range. Transmittance, $T = 1-R$, is calculated by the Fresnel reflection coefficient, $R = [({n_1-n_2})/({n_1+n_2})]^2$, which specifies transmission and reflection at a perfectly flat interface between two transparent homogeneous media. The absorbance is neglected in this model. 

For the geometric focal point, the optical path should be regulated based on Snell's law \cite{Coua07, Boyd08},
\begin{equation}
n_{1}\sin\theta_{1}=n_{2}\sin\theta_{2}.\\    
\end{equation}
The $n_1$ and $n_2$ are incident and refracted indices of the air and the 4H-SiC, and the {$\sin{\theta_1}$} and {$\sin{\theta_2}$} are incident and refracted angles. According to Snell's law, the maximum incident angle of the beam was determined by the NA of the objective lens. Given the refractive index of 4H-SiC, the corresponding refracted angle enables control of the geometric focal depth inside the medium as in Fig. 2b. 
Here $z_f$ denotes the nominal linear-optics focus inside 4H-SiC, set by the objective $z$-translation with a refractive index correction. Because 4H-SiC exhibits a positive Kerr coefficient ($n_\mathrm{Kerr} > 0$), the on-axis refractive index increases with intensity, acting as a self-induced lens. As the linearly focused beam approaches $z_f$, the on-axis intensity rises, and the collapse criterion is met upstream of the geometric focus, yielding $z_{\mathrm{sf}} < z_f$. The collapse is then arrested by plasma-induced defocusing and nonlinear absorption $\cite{Coua07, Berge07}$.

\subsection{Prediction of self-focusing depth using a modified Marburger model}
\label{2.3}
The self-focusing point has been estimated in the semi-empirical equations by Marburger \cite{Marb75}, for a collimated beam configuration. In laser slicing using a focused beam, we approximated the parameters to the spatial rate of radius change far from the beam waist, as described in Appendix B, for practical applications. The prediction of the self-focusing depth was performed using a modified Marburger equation,
\begin{equation}
z_\mathrm{sf}/z_\mathrm{f} = \frac{0.367}{\sqrt{\left[ \sqrt{I_{\mathrm{in}} \left/ I_{\mathrm{cr}} \right.} - 0.852 \right]^2 - 0.0219}}.
\end{equation}
Here, $I_\mathrm{in}$ denotes the on-axis peak irradiance of the incident beam at the geometrical focal plane in the absence of nonlinear effects. The term $I_\mathrm{cr}$ is used as an effective critical irradiance that normalizes the nonlinear strength, which is defined from the experimentally measured surface damage threshold.

In the modified Marburger equation, the $z_\mathrm{f}$ represents the location of beam collapse at the geometrical focus, reflecting the actual focusing condition. This modification accounts for the tight focusing by an objective lens, where Kerr-effect-induced self-focusing leads to a shift in the effective focal position. For tightly focused beams, the modified form provides a more accurate estimation of the nonlinear focus location. In addition, the denominator representing the nonlinear strength has been reformulated. Instead of the original power ratio $P_\mathrm{in}/P_\mathrm{cr}$, we employed the irradiance ratio $I_\mathrm{in}/I_\mathrm{cr}$, which more accurately characterizes the nonlinear response near the focal region. 
The $I_\mathrm{cr}$ value is not identical to the true Kerr critical power $P_{\mathrm{cr}}$, but when used in the form of an irradiance ratio $I_{\mathrm{in}}/I_{\mathrm{cr}}$, it preserves the correct monotonic dependence of the nonlinear focusing strength. Although the absolute physical meaning differs, the trend governing self-focusing behavior remains consistent within our model.
In practical terms, the local irradiance is influenced by the rapidly varying beam cross-sectional area along the propagation of the beam axis, and it plays a critical role in initiating plasma generation. Therefore, adopting the irradiance ratio in the modified equation allows for a more precise representation of the spatially localized nature of the nonlinear interaction.

\subsection{Wafer separation behavior and post-slicing characterization}
\label{2.4}
Separation stress was measured during the wafer separation to evaluate the processability of the laser slicing using the universal testing machine (Multitest1-i, Mecmesin, U.K.). An adhesive on the as-sliced wafer surface was glued on the tip of the tensile test machine. The available maximum load to sample is 950 N. Surface profiles were analyzed by a confocal microscope (VK-X200K, Keyence, Japan) with a 1 $\mu$m display resolution. Crystalline peaks are analyzed by the Raman spectrometer (Heda 250, Weve Inc., Korea).

\subsection{Nonlinear beam propagation modeling and simulation}
\label{2.5}
We simulated femtosecond laser beam propagation in 4H-SiC using a Kerr-type refractive index model in which the local index increases with intensity. The details of the simulation modeling and equations are described in Appendix C. Material parameters correspond to 4H-SiC at a wavelength of 1040 nm. The incident pulse energy was 2 to 6 $\mu$J with a duration of 400 fs. A surface reflectivity of 0.19 was applied to obtain the transmitted energy. A paraxial Gaussian beam was focused to a prescribed geometric depth, with a waist diameter of 2.746 $\mu$m, and the linear depth-dependent Gaussian profile was used to seed the nonlinear index before ray tracing. Fields were discretized on a Cartesian grid spanning 50 $\mu$m laterally and 500 $\mu$m in depth resolution. Ray propagation followed a geometric-optics formulation. We launched 500 rays from the entrance plane with uniformly spaced lateral positions, integrated their trajectories with a fixed arclength step while enforcing unit speed, and terminated rays that exited the domain.

Spatiotemporal intensity was reconstructed in an energy-conserving manner by assigning each ray a Gaussian launch weight, distributing its transmitted energy uniformly along the in-bounds samples of its path, and depositing that energy onto the grid with a compact smoothing kernel. The total deposited energy matched the transmitted pulse within numerical precision. From the reconstructed intensity map we extracted: (i) the first self-focusing depth, defined as the axial location of the on-axis peak; (ii) the lateral spot size at that depth, reported as the full width at half maximum from the transverse profile; and (iii) the fraction of pulse energy contained in an affected region, defined by a 5\% of the peak intensity. Unless otherwise stated, absorption and plasma hydrodynamics were not included. All observed data were computed directly from the reconstructed fields.

\section{Results and Discussions}
\label{sec3}
\subsection{Critical irradiance and processability window in self-focusing driven 4H-SiC slicing}
\label{3.1}
In the modified Marburger equation, $z_{\mathrm{sf}}$ is governed by an irradiance threshold for 4H-SiC slicing. 
In the process, we adopt the $I_{\mathrm{cr}}$ for the surface damage threshold (SDT) as the normalization and as the background of the map. By contrast, the onset of bulk plasma-assisted lattice collapse does not occur at a single well-defined irradiance, but varies significantly with depth, propagation distance, and path-dependent nonlinear losses such as multiphoton absorption and free carrier absorption. Because the bulk threshold is not spatially uniform and changes along the beam trajectory, 
it cannot serve as a reliable single normalization constant for constructing 
a process map. For this reason, we use the experimentally measured surface damage threshold as $I_{\mathrm{cr}}$, which provides a stable and reproducible reference irradiance that is independent of subsurface heterogeneity and optical interference effects. Although the absolute bulk threshold is higher than the surface threshold, using the surface value in the dimensionless ratio $I_{\mathrm{in}}/I_{\mathrm{cr}}$ preserves the correct monotonic dependence of the nonlinear response and leads to consistent agreement between the model predictions and experimental observations. This choice offers high measurement reproducibility, directly corresponds to the SDT, and provides a conservative normalization base.

\begin{figure}[h!]
\centering
\includegraphics[width=\linewidth]{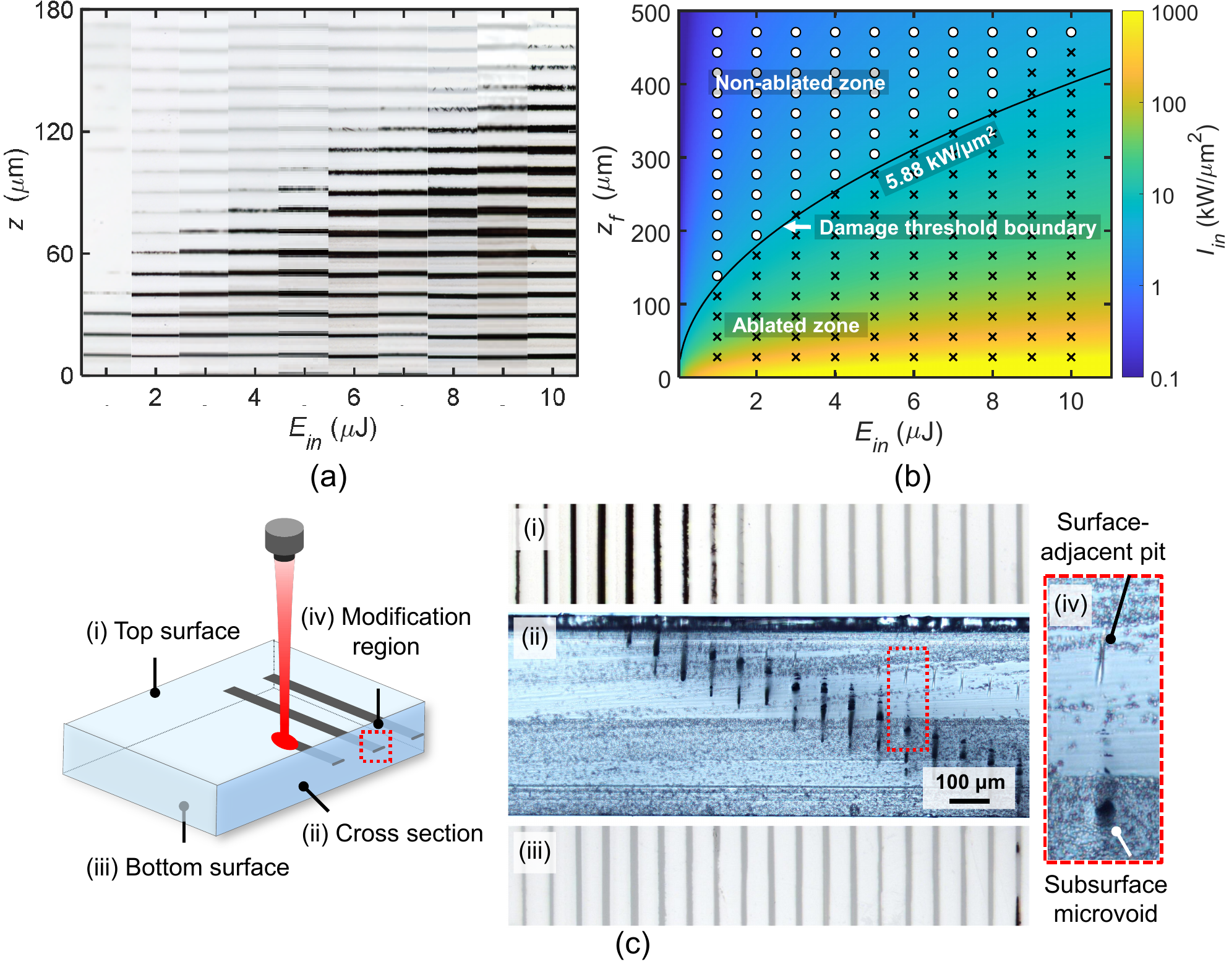}
\caption{(a) Images of the laser beam-ablated regions on the SiC wafer surface, captured while varying the pulse energy, decreasing the distance between the objective lens and the wafer surface ($z$) in 10 $\mu$m steps from the focal position. (b) Colormap showing the irradiance at the wafer surface as a function of incident pulse energy ($E_\mathrm{in}$) and geometric focal depth ($z_\mathrm{f}$). Symbols indicate the investigated regions: circles (o) represent surfaces without ablation damage, while crosses (×) indicate the ablated surfaces.
(c) Evolution of the microvoid location in cross section while translating the objective lens along $z$ at an incident pulse energy $E_\mathrm{in}$ = 2 $\mu$J. Microscopic images show (i) the top surface, (ii) the cross-section of the processed track, and (iii) the bottom surface. (iv) A magnified view of the modification region highlighting a surface-adjacent pit near the air/4H-SiC interface and subsurface microvoid(s) within the laser-modified layer.}
\label{fig3}
\end{figure}

To examine the ablation tendency, we systematically varied the surface irradiance by increasing the pulse energy from 1 to 10 $\mu$J in 1 $\mu$J increments and by adjusting the geometric focal depth from 0 to 180 $\mu$m in 10 $\mu$m steps, as illustrated in Fig. 3a.
The surface ablation results are plotted with symbols on the color map, which describes the calculated incident irradiation at the wafer surface level in Fig. 3b; the circles (o) represent no ablation occurred, meaning a clean surface, and the crosses (×) indicate surface ablation occurred. The boundary of the ablation threshold appears along the solid line with the 5.88 kW/$\mu m^2$ of the incident irradiation. A damage threshold boundary with a constant means that a focused beam could be dealt with regarding irradiance to describe a critical point by the femtosecond laser modification. Therefore, we adopt $I_{\mathrm{cr}}= 5.88$ kW/$\mu m^2$ for normalization in the processability map and subsequent analysis. 
It should be noted that this threshold value is not universal. The effective surface damage irradiance can vary depending on wafer surface condition, crystal quality, laser beam quality, and objective-lens aberrations. In here, all optical and material parameters other than incident pulse energy and geometric focal depth were held fixed. This ensures that $I_{\mathrm{cr}}$ serves as a consistent internal reference for comparing different slicing conditions within the same optical system.

Fig. 3c visualizes how the microvoid position evolves with the objective $z$-translation at an incident pulse energy of $E_{\mathrm{in}} = 2~\mu\mathrm{J}$. To enable direct cross-sectional observation, a 4H-SiC specimen was pre-cleaved, and machining was initiated from outside the exposed cross-section. Single modification tracks traverse vertically across the cross-section, forming a well-defined internal modification region. 
In Fig. 3c(ii), the dark, needle-like modification traces observed along the beam path indicate localized energy collapse induced by Kerr-driven self-focusing inside the crystal \cite{Jian24}.
In addition to subsurface microvoids in Fig. 3c(iv), a surface-adjacent pit is visible at the entry surface of the cross-section. The feature corresponds to the focal imprint formed near the air/4H-SiC interface before the beam propagates into the bulk. Subsurface microvoids then develop progressively deeper inside the crystal, where the higher refractive index enhances beam convergence, leaving a continuous depthwise trace along the propagation direction. 
As the incident irradiance approaches the nonlinear threshold, self-focusing compresses the beam core and sustains a high-intensity channel, resulting in confined energy deposition and filament-like internal modification rather than uniform volumetric damage \cite{Wang22-2}.
The systematic shift of both surface-adjacent pits and internal modification features with objective $z$-translation confirms the consistent control of the focal geometry.

For completeness, the high refractive index can also result in a longitudinal elongation of the focal region rather than a rigid shift of the focal point \cite{Yuan23}.
The wafer surface was used as the reference plane for defining the geometric focal position. This procedure does not constitute aberration correction but merely establishes a consistent spatial origin for comparing processing conditions. While spherical aberration influences the absolute axial distribution of irradiance, its contribution remains unchanged in form for all experiments performed under fixed optical conditions (objective lens, NA, and wavelength).
As quantified in Appendix D using classical aberration theory, the longitudinal focal elongation induced by refractive index mismatch under the present experimental conditions is limited to only a few micrometers, even when the geometric focus is positioned at the bottom of the 4H-SiC surface. Consequently, the relative variation of the modification depth with incident irradiance, attributed to Kerr-induced self-focusing, can be isolated and analyzed independently of the static aberration background.

\subsection{Experimental decoupling of pulse overlap and inter-pulse temporal spacing}
\label{3.2}
During the formation of a modification track, the modification depth is governed by Kerr-induced self-focusing. Also, the modification depth is further modulated by the pulse overlap condition, which is controlled by the scanning speed and repetition rate. Changes in pulse overlap can shift the effective self-focusing depth and the resulting multifocal modification pattern, either through cumulative modifications of the optical path or through thermal effects associated with high repetition rate irradiation. To decouple these contributions, we separately investigated the scanning speed-driven pulse overlap variations at fixed average power and repetition rate and the inter-pulse temporal spacing
at constant pulse energy with proportionally adjusted scanning speed. The notation and statistical descriptors used to quantify the depths of the observed modification features are summarized in Appendix E.

The modification depth was described as the variation under identical pulse energy while changing only the scanning speed. Fig. 4a-c shows the depth distribution of microvoids produced by 10 consecutive single track scans at a fixed geometrical focus depth of $z_\mathrm{f}$ = 343 $\mu$m. The magnified cross-sectional views in Fig. 4a1-c1 reveal a clear upward shift of the incident-side boundary of the primary modification depth, $z_\mathrm{m}$, accompanied by an expansion of the principal modification distribution, $s_{\bar{z}_{\mathrm{m},p}}$, along the depth direction. 
As summarized in Fig. 4d, decreasing the scanning speed increases the pulse overlap fraction. For a focal beam spot diameter of 2.746 $\mu$m, scanning speeds of 150, 75, and 37.5 mm/s yield area overlap fractions of 65.6\%, 82.6\%, and 91.3\%, respectively. 
Under the same pulse energy, the higher overlap shifts the formation of the primary modification point toward the incident surface. This trend is consistent with prior reports on overlap-driven cumulative effects that promote an earlier onset of modification along the beam path \cite{Zhan23}. 
The centroid depth of the principal modification distribution, $\bar{z}_{\mathrm{m},p}$, calculated from principal modification points with a feature size >5 $\mu$m, indicates that the multifocal modification region not only shifts but also becomes broader and extends further in depth as the overlap increases.

\begin{figure}[h!]
\centering
\includegraphics[scale=0.55]{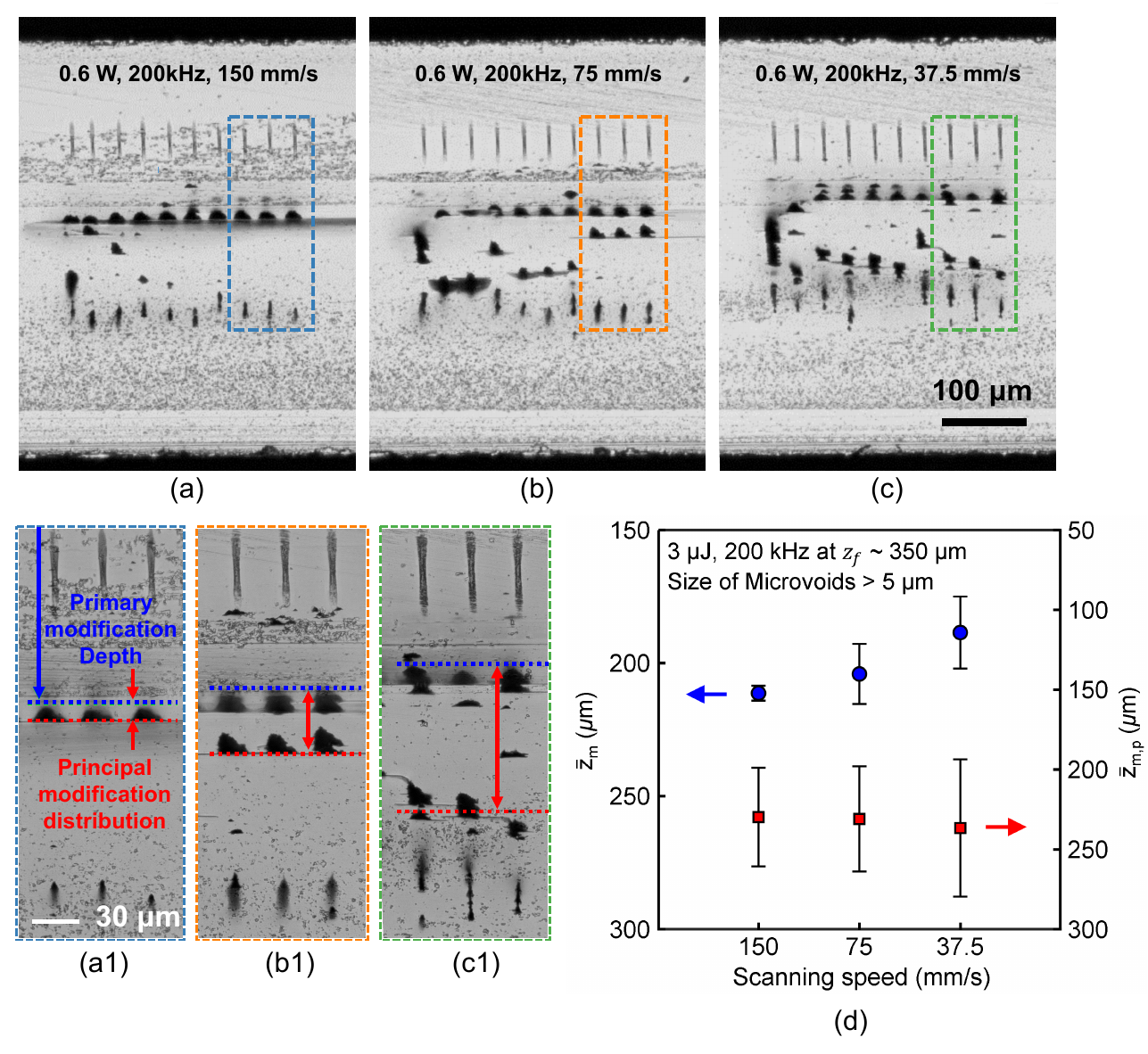}
\caption{Effect of pulse overlap on modification region in self-focusing-driven 4H-SiC slicing. The average laser power (0.6 W) and repetition rate (200 kHz) were fixed, while only the scanning speed was varied: (a) 150 mm/s, (b) 75 mm/s, and (c) 37.5 mm/s. The corresponding magnified views (a1-c1) highlight the primary modification depth, $z_\mathrm{m}$, and the set of principal modification sites distributed along the depth direction. (d) Modification depth as a function of scanning speed showing $\bar{z}_\mathrm{m}$ and the mean depth of the principal modification distribution, $\bar{z}_\mathrm{m,p}$. Data points indicate the mean, and error bars denote $\pm 1$ standard deviation (SD), across 10 independent tracks (n=10).}
\label{fig1}
\end{figure}

Fig. 5 examines the characteristics of the principal modification distribution and its first self-focusing collapse under a constant pulse overlap condition while varying the temporal accumulation scale. To maintain the same area overlap fraction, the reference condition (0.6 W, 200 kHz) was scaled such that the pulse energy remained fixed at 3$\mu$J. Specifically, the average power and scanning speed were reduced in proportion to the decrease in repetition rate (0.6 W, 200 kHz, 75 mm/s; 0.3 W, 100 kHz, 37.5 mm/s; and 0.15 W, 50 kHz, 18.75 mm/s). With this scaling, the number of pulses delivered per unit area was preserved, while only the pulse-to-pulse temporal spacing was altered, as illustrated by the comparable cross-sectional morphologies in Fig. 5a1-c1.
Despite the longer inter-pulse time at lower repetition rates, the microvoid distribution and the principal modification distribution remained qualitatively similar. Correspondingly, Fig. 5d shows that the primary modification depth, $z_\mathrm{m}$, exhibits a systematic shift with $z_\mathrm{f}$ but no pronounced dependence on the scanning speed and repetition rate scaling under constant overlap fraction.

\begin{figure}[h!]
\centering
\includegraphics[scale=0.4]{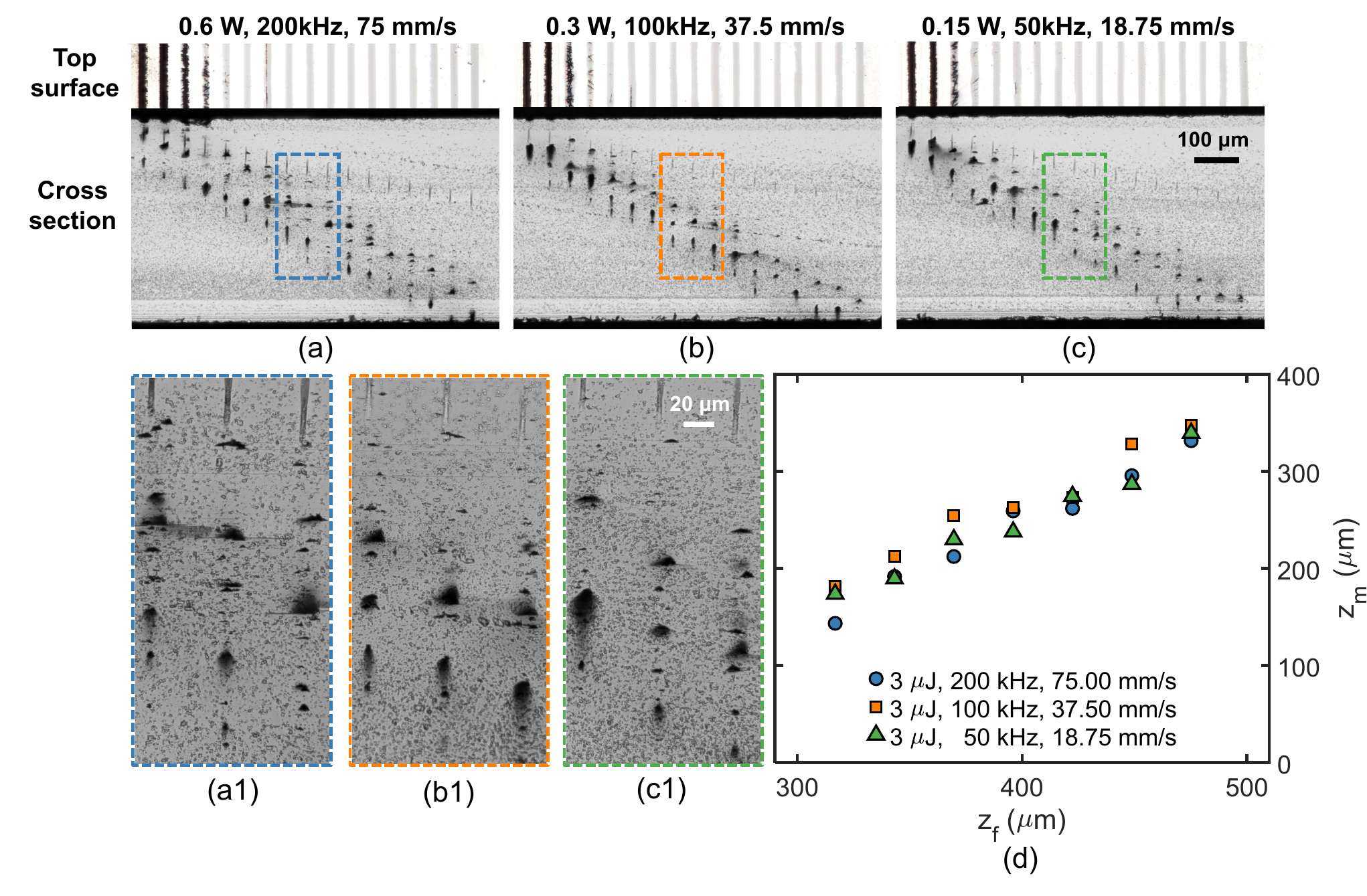}
\caption{Effect of inter-pulse temporal spacing at constant pulse energy on the modification region in self-focusing-driven 4H-SiC slicing. The pulse energy was fixed at 3 µJ by proportionally scaling the average power, repetition rate, and scanning speed: (a) 0.6 W, 200 kHz, 75 mm/s; (b) 0.3 W, 100 kHz, 37.5 mm/s; and (c) 0.15 W, 50 kHz, 18.75 mm/s. 
Magnified cross-sectional views at the same processing location are shown in (a1-c1). By varying the geometrical focus depth, $z_\mathrm{f}$, (d) summarizes the representative primary modification depth, $z_\mathrm{m}$, extracted from the principal modification region with feature size $>$5 $\mu$m for each $z_\mathrm{f}$ condition.}
\label{fig1}
\end{figure}

\begin{figure}[h!]
\centering
\includegraphics[scale=0.55]{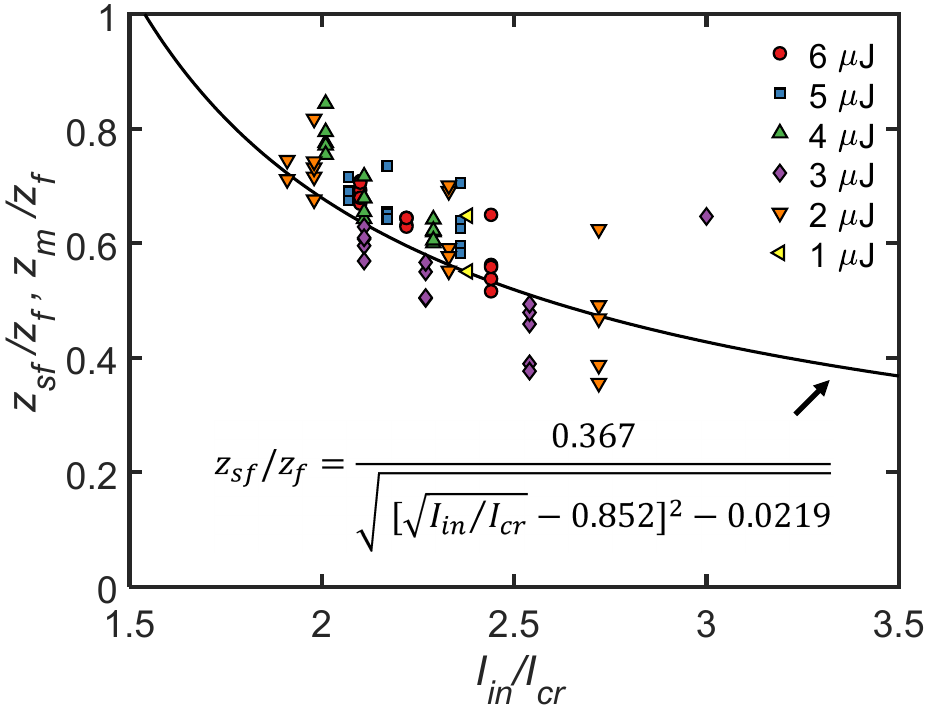}
\caption{Relationship between the self-focusing and the geometrical focus as a function of normalized irradiance ($I_{\mathrm{in}}/I_{\mathrm{cr}}$). The solid line represents the modified Marburger equation (the formula indicated inside the graph), and all symbols indicate experimental results, showing the normalized modification depth ($z_{\mathrm{m}}/z_{\mathrm{f}}$) versus normalized irradiance.}
\label{fig4}
\end{figure}

Experimental points for the modification depth ratio ($z_{\mathrm{m}}/z_{\mathrm{f}}$) versus the normalized irradiance ($I_\mathrm{in}/I_\mathrm{cr}$) are plotted together with an analytical curve from the modified Marburger formula in Fig. 6. Data were obtained from wafer slicing experiments at pulse energies $E_\mathrm{in}$=1 to 6 $\mu$J. For each irradiation condition, five independent trials were attempted at each of the three to four specified geometric focal depths, using separately prepared 4H-SiC wafers to ensure statistical independence. Only the separable samples with a non-ablated top surface were included in the quantitative analysis. Some samples based on the slicing conditions did not separate even under the maximum applied loading (e.g., $E_\mathrm{in}$=1 $\mu$J); in such cases, the separation stress could not be defined, and these failed attempts were excluded from the statistical evaluation.

Processable range is limited to around 2 to 3 of the $I_\mathrm{in}/I_\mathrm{cr}$. The geometric focal depth denotes the virtual focus inside the medium, the axial location set solely by of refractive index $n_\mathrm{0}$, where the focused beam would converge without light-matter interaction in Fig. 2b. For each experimental point, $I_\mathrm{in}$ = $P_\mathrm{in}$/[$\pi w_\mathrm{lin} (z_\mathrm{sf})^{2}$] was calculated at the model-predicted collapse position using the pre-collapse radius $w_\mathrm{lin}(z)$ to avoid a circular dependence between the $I_\mathrm{in}$ and $z_\mathrm{sf}$. The details of the $w_\mathrm{lin}$ are described in Appendix A. Repeated measurements at identical $E_\mathrm{in}$, and $z_\mathrm{f}$ in the experimental points around the model curve. Within the operating window, only the first collapse is considered as the effective processing point. In the analytical model, the self-focusing effect requires an irradiance exceeding at least 1.54 times the critical irradiance. A decreasing modification depth ratio indicates an increasingly nonlinear response as the irradiance ratio increases. In particular, the irradiance ratio increases as the geometric focus approaches the wafer surface.

\subsection{Numerical analysis of Kerr-induced self-focusing and collapse behavior}
\label{3.3}

\begin{figure}[h!]
\centering
\includegraphics[width=\linewidth]{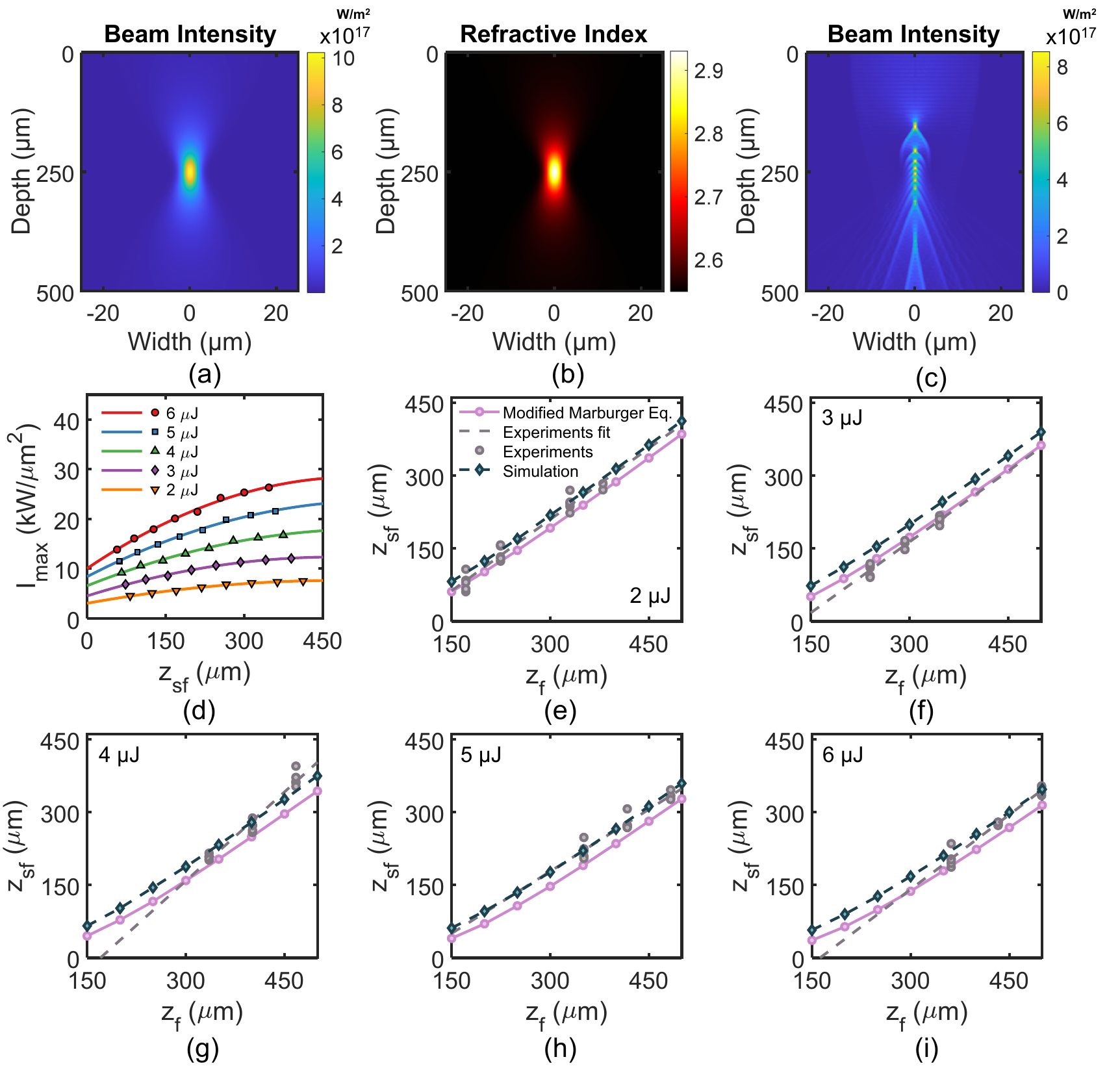}
\caption{Results of the ray optics simulation of a femtosecond laser beam path in the 4H-SiC. Simulation results of (a) the beam intensity profile without the Kerr effect, (b) the refractive index profile with the Kerr effect, and (c) the beam intensity profile with Kerr effect, all at incident pulse energy $E_\mathrm{in}$ = 3 $\mu$J and geometric focal depth $z_\mathrm{f}$ = 250 $\mu$m. (d) Simulation results of the maximum beam irradiance at the self-focusing depth. (e-i) Self-focusing depth for $E_\mathrm{in}$ = 2 to 6 $\mu$J, respectively,  from simulations using the nonlinear refractive index $n_\mathrm{Kerr}$ of 4H-SiC, from the analytical model, and from the experimental data with a fit curve.}
\label{fig5}
\end{figure}

Ray optics simulation results are compared with the analytical model and with experimental data. The nonlinear refractive index significantly changes the evolution of the beam intensity in the femtosecond laser slicing. In Fig. 7a-c, the beam intensity profiles and refractive index field are described at the 3 $\mu$J of input pulse energy with the 250 $\mu$m of geometrical focal depth. Without a nonlinear refractive index, the focal point coincides with the geometric focus (Fig. 7a). Considering the Kerr effect (Fig. 7b), the refractive index varies with intensity, which bends rays toward the axis and advances the effective focus. 

In the simulations, the beam collapses upstream of the geometric focus. Then it alternates between diffraction-driven defocusing and refocusing, producing multiple irradiance peaks on-axis along the propagation direction (Fig. 7c). The multiple refocusing behavior is expected when the peak power exceeds the critical power. The presented ray trajectory simulation neglected free electron plasma generation and absorption. It should be emphasized that the purpose of the simulation is not to reproduce the full transient laser-matter interaction after plasma or permanent damage formation. Instead, the model is intentionally simplified to capture the onset and spatial location of the first Kerr-induced self-focusing collapse, which governs the initial energy localization inside the crystal. The simulation visualizes how the total refractive index, modified by the nonlinear refractive index term, reshapes the beam intensity distribution prior to significant material transformation. The filament-like features along the axial beam path should be interpreted as repeated Kerr-induced self-focusing and defocusing in an idealized, undamaged medium, rather than as a direct representation of the subsequent material response observed experimentally.

In practical femtosecond laser focusing, the first nonlinear collapse induces a localized refractive index modification on the beam axis, which effectively defocuses subsequent propagation. Concurrently, photo-excited free carriers introduce collisional damping through electron-phonon interactions and defect scattering, leading to increased optical attenuation \cite{Coua07}. The deposited energy initially raises the electron temperature, followed by lattice heating through electron-phonon coupling on a picosecond timescale. This plasma-mediated energy transfer promotes localized material modification, resulting in enhanced scattering from microvoids, cracks, and defect structures, which further suppresses downstream refocusing to the following repetitive beams.

Only the first self-focusing point is considered as the effective processing region in practice. Although multi-focusing appears in the simulations, downstream refocusing traces were not dominant in the experiments. Fig. 7d summarizes the on-axis peak irradiance ($I_\mathrm{max}$) at collapse and its location. Simulation data points were sampled from 150 to 500 $\mu$m in 50 $\mu$m steps together with a fit curve to map the dependence on focus position. For a fixed pulse energy, deeper geometric focusing leads to larger $I_\mathrm{max}$ at $z_\mathrm{sf}$. This trend is consistent with the combined action of linear focusing and the cumulative Kerr effect before collapse \cite{Coua07}. The increase of $I_\mathrm{max}$ with depth is more pronounced at higher pulse energy. In all cases, the collapse occurred upstream of the geometric focus. Moreover, as $E_\mathrm{in}$ increases, the collapse shifts earlier, reflecting the stronger Kerr nonlinearity at higher pulse energy. Practically, the beam intensity profiles should be considered as an upper bound on the extent of multi-focusing behavior, while the first self-focusing depth remains a robust point of comparison with experiment. The location of the first self-focusing depth shows a similar slope with respect to the geometric focus offset in both the experiments and the analytical model at input pulse energies of 2 to 6 $\mu$J, respectively, in Fig. 7e-i. The results agree within experimental uncertainty.

\begin{figure}[h]
\centering
\includegraphics[scale=0.53]{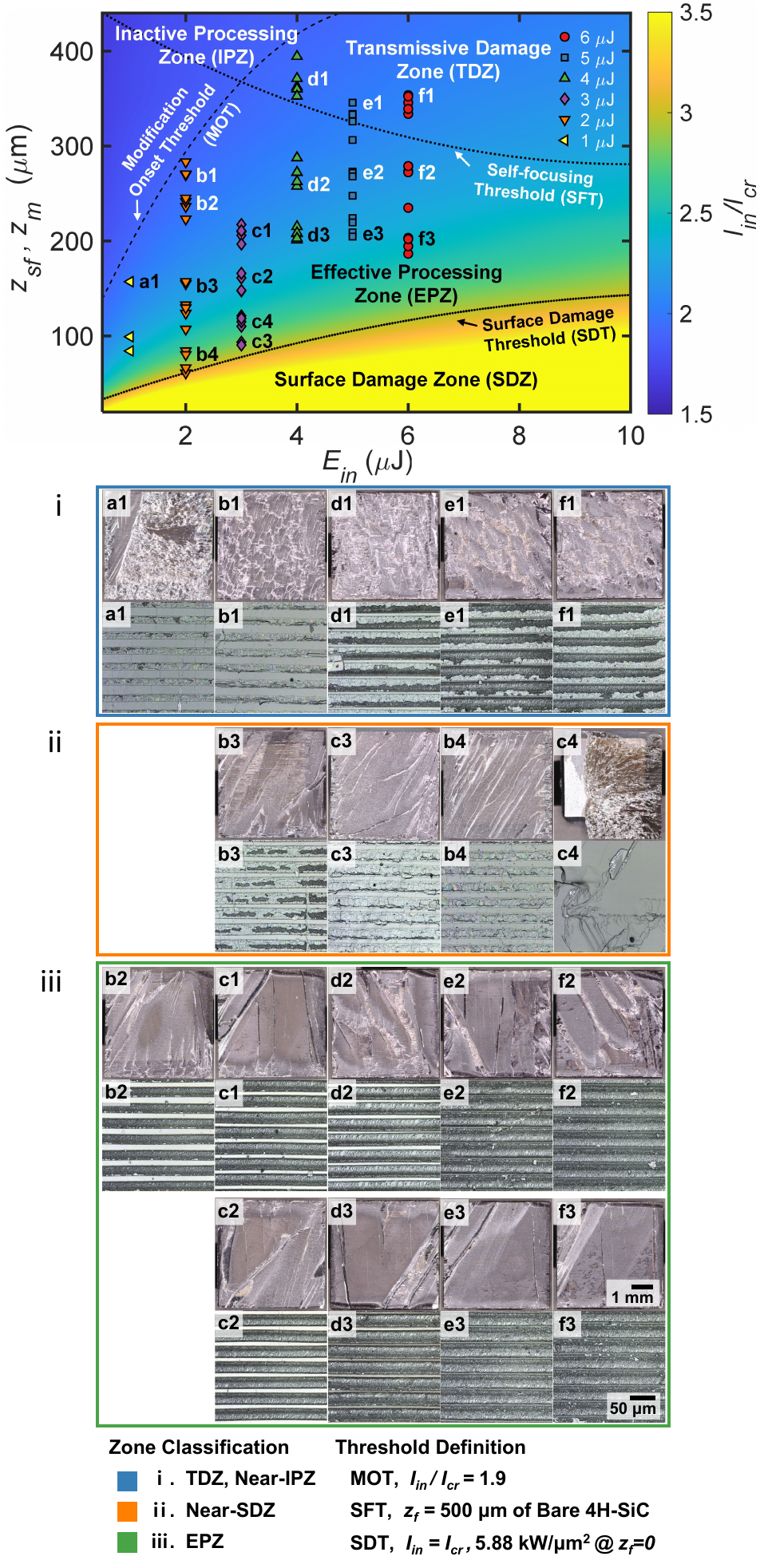}
\caption{Processability map for laser slicing with the three theoretical thresholds. The colormap represents the normalized irradiance at the self-focusing depth. Optical microscope images at both low and high magnification show the surface texture of the sliced wafer under the corresponding conditions. Each image corresponds to the labeled notations within the colormap.}
\label{fig6}
\end{figure}

\clearpage

\begin{table}[h!]
\centering
\small
\caption{Separation stress ($\sigma$) and surface texture parameters ($S_\mathrm{a}$, $S_\mathrm{q}$, $S_\mathrm{sk}$, $S_\mathrm{ku}$) for samples grouped according to the processability classification in Fig. 8.}
\vspace{2mm}

\begin{tabular}{lllllll}
\toprule
\textbf{\makecell{Group}} &
\textbf{\makecell{Notation}} &
\textbf{\makecell{Separation \\ Stress \\ $\sigma$ (MPa)}} &
\textbf{\makecell{Areal \\ Average \\ Surface \\ Roughness \\ $S_\mathrm{a}$ ($\mu$m)}} &
\textbf{\makecell{Areal \\ RMS \\ Surface \\ Roughness \\ $S_\mathrm{q}$ ($\mu$m)}} &
\textbf{\makecell{Surface \\ Height \\ Skewness \\ $S_\mathrm{sk}$}} &
\textbf{\makecell{Surface \\ Height \\ Kurtosis \\ $S_\mathrm{ku}$}} \\
\midrule

\textbf{i}               & a1 & 28.98 & 18.43 & 23.49 & -0.17 & 3.69 \\
                         & b1 & 26.04 & 22.00 & 27.24 & -0.87 & 3.57 \\
                         & d1 & 24.02 & 20.00 & 28.56 & -1.97 & 7.36 \\
                         & e1 & 17.28 & 28.96 & 41.18 & -1.83 & 5.52 \\
                         & f1 & 18.20 & 34.07 & 44.81 & -1.56 & 4.31 \\
                         
\textbf{ii}              & b3 & 12.85 & 14.26 & 17.03 & 0.49 & 2.46 \\
                         & b4 & 20.30 & 8.60 & 10.58 & 0.30 & 2.92 \\
                         & c3 & 26.55 & 6.46 & 8.67 & -0.41 & 4.96 \\
                         & c4 & 18.29 & 17.01 & 20.69 & 0.38 & 2.41 \\

\textbf{iii}             & b2 & 3.68 & 18.84 & 23.45 & 0.10 & 2.90 \\
                         & c1 & 1.12 & 8.69 & 11.14 & -0.23 & 4.38 \\
                         & c2 & 6.09 & 12.39 & 17.08 & -1.42 & 6.97 \\
                         & d2 & 1.92 & 10.68 & 17.11 & -2.85 & 18.02 \\
                         & d3 & 0.94 & 9.29 & 19.29 & -3.94 & 21.12 \\
                         & e2 & 5.42 & 8.13 & 15.84 & -4.15 & 26.87 \\
                         & e3 & 1.12 & 5.30 & 6.42 & -0.27 & 3.61 \\
                         & f2 & 3.85 & 13.64 & 21.71 & -2.59 & 16.29 \\
                         & f3 & 2.72 & 5.29 & 8.09 & -1.40 & 7.37 \\
\bottomrule
\end{tabular}
\end{table}

\subsection{Self-focusing controlled processability and effective processing zone}
\label{3.4}
All of the separable samples for the 4H-SiC wafer slicing conditions are described in Fig. 8. High- and low-magnification microscopic image pairs for the representative conditions are depicted for each condition by the energy pulse and the self-focusing depth. The image pairs are classified into three color-coded box lines in sequence: blue (near-inactive), orange (near-damaged), and green (proper). The processing map depicts the EPZ within the boundaries of the SDT, the self-focusing threshold (SFT), and the modification onset threshold (MOT). The thresholds set three boundaries to the pulse energy with the modification and the self-focusing depths on the colored map of the irradiance ratio. 

The SDT boundary corresponds to the experimentally measured irradiance at which surface ablation and diffuse reflection begin to deteriorate the slicing quality. The SFT boundary is set by the refracted geometric focal depth exceeding the wafer thickness. Beyond the SFT, the beam collapses outside the wafer and may damage the surrounding medium, even though Kerr-induced self-focusing can still occur upstream inside the material. The MOT boundary was determined from the minimum irradiance ratio required to form a continuous internal modification track while keeping the top surface intact. In our experiments, the condition $I_{\mathrm{in}}/I_{\mathrm{cr}} \ge 1.9$ consistently produced separable samples within the measurable separation-stress range. The MOT value is not universal, as it may vary depending on different scan speeds, line spacings, or sample geometries. The MOT serves as a practical boundary of the EPZ under the fixed slicing parameters used. With these three boundaries, the processability map provides a physically grounded segmentation of the slicing response, distinguishing the proper EPZ from the near-inactive and near-damaged regimes.

Surface profiles at the modification region on the optical microscope images could be classified into three groups based on the classification as shown in Fig. 8. 
It should also be noted that the fracture morphology of 4H-SiC can depend on the crystallographic orientation. In our previous research \cite{Youn25}, the separation-induced crack paths tend to follow specific crystal directions, meaning that the appearance of surface features in Fig. 8 may vary slightly with in-plane scanning orientation or Si-/C-face differences, even under identical irradiance conditions. 

In the first image group i, the TDZ and near-IPZ samples (blue) exhibit a distinct mesh-like pattern across the entire surface, which is clearly visible in the low-magnification optical images. At higher magnification, the detailed morphology becomes apparent. The wide gaps between beam paths for a1 and b1 indicate the MOT-proximal regime, whereas the images d1, e1, and f1 show light-gray SiC fragments remaining on the dark beam path region, revealing partially detached micro-debris along the cut trajectory.
Second image group ii, near-SDZ (orange) on the higher magnification images, describes the fragments remaining on the beam path region. Even the beam path trace could not be found on the separated surface in the c4 image. Last image group iii, EPZ (green), has a dark gray region with diagonal crack patterns. The beam path region is clear without any fragility in the low magnification images, which has a dominant dark gray region on the higher magnifications.  

Table 1 summarizes the separation results along with the separation stress ($\sigma$) and surface texture parameters ($S_\mathrm{a}$, $S_\mathrm{q}$, $S_\mathrm{sk}$, $S_\mathrm{ku}$). The groups (i-iii) were defined from the processability map in Fig. 8 by clustering conditions that exhibit similar macroscopic slicing response. As expected, the $\sigma$ and $S_\mathrm{a}$ show clear grouping: group iii exhibits the lowest $\sigma$ (typically $<10$ MPa) and relatively small $S_\mathrm{a}$ ($<20$ $\mu$m), whereas group ii shows higher $\sigma$ despite $S_\mathrm{a}$ remaining comparable to group iii. In addition, group i shows higher $\sigma$ and $S_\mathrm{a}$ than group iii. This grouping is consistent with the processability map and indicates that the suitability of slicing could be separated distinctly on the $\sigma$-$S_\mathrm{a}$ plane. Consistent with the previous microscopic-feature analysis, $S_\mathrm{q}$ follows essentially the same trend as $S_\mathrm{a}$. Both parameters become larger in groups i and ii, where the separation stress is also high, confirming that the overall amplitude of the topography correlates with the required separation force.

The skewness values $S_\mathrm{sk}$ are predominantly negative across all groups, indicating valley-dominated height distributions rather than peak-dominated ones. Such valley-type morphology suggests that femtosecond slicing mainly produces crack-like depressions, and the process terminates before extensive melt resolidification can generate pronounced recast ridges on the cut surface. The kurtosis $S_\mathrm{ku}$ provides additional information about the microvoid structures, particularly for group iii, where the self-focusing collapse occurs near the center of the EPZ. In these conditions, $S_\mathrm{ku}$ increases markedly while $S_\mathrm{a}$ and $S_\mathrm{q}$ remain relatively low and $|S_\mathrm{sk}|$ becomes large. The low amplitude $S_\mathrm{a}$ but high skewness magnitude and kurtosis indicate the presence of frequent, sharp crack-like valleys of limited lateral extent. Such finely distributed microvoids promote efficient wafer separation while maintaining comparatively smooth surfaces, consistent with the substantially reduced separation stress observed in group iii. In summary, the texture parameters confirm that optimal EPZ slicing produces a dense population of narrow valley type features rather than large scale grooves, thereby explaining both the low $\sigma$ and modest surface roughness observed in group iii.

\subsection{Microscopic surface morphology along the beam path region}
\label{3.5}

\begin{figure}[h!]
\centering
\includegraphics[scale=0.5]{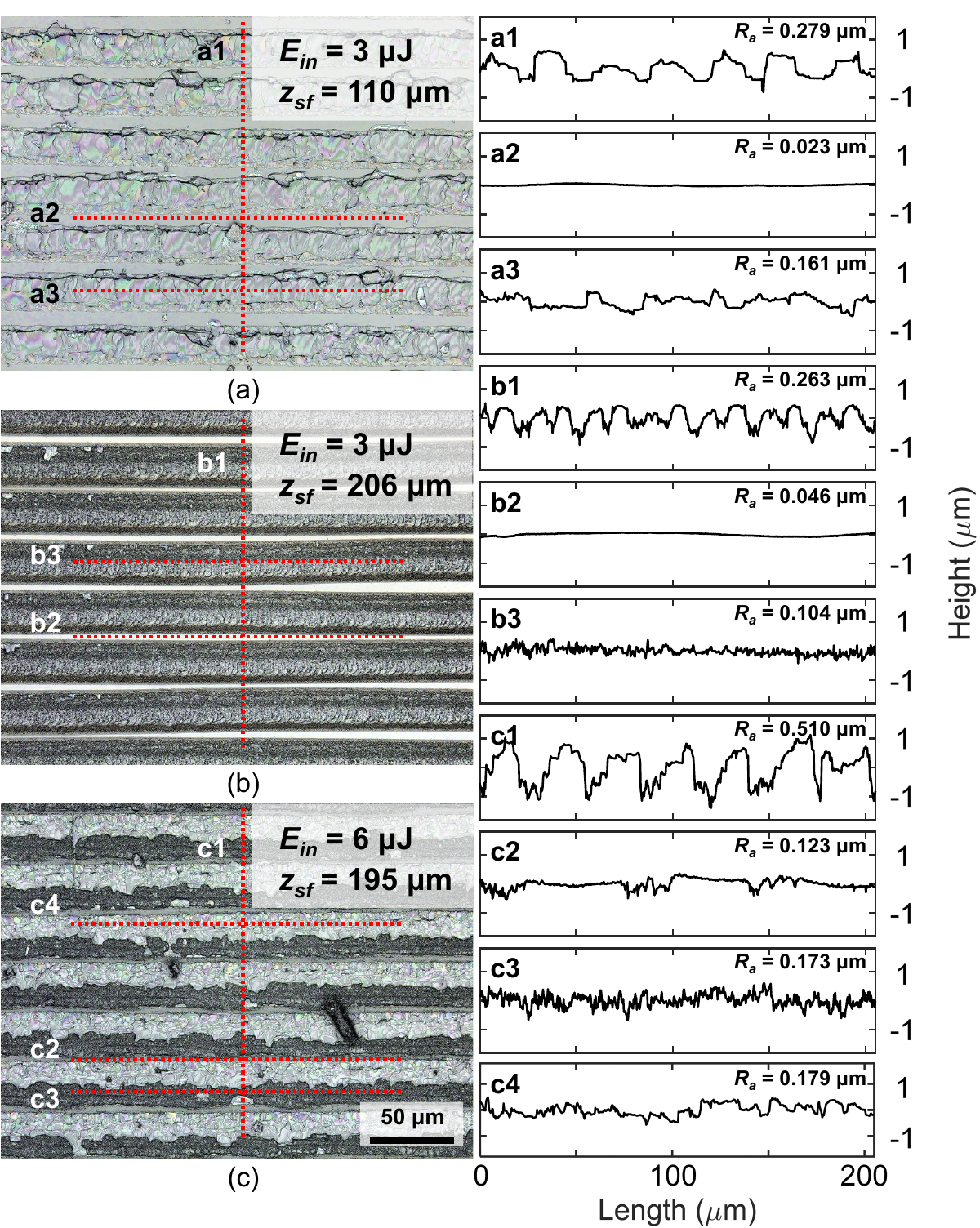}
\caption{Optical microscopic images of the sliced 4H-SiC wafer surface under three representative processing conditions: (a) $E_{\mathrm{in}} = 3~\mu$J and $z_{\mathrm{sf}} \approx 100~\mu$m, (b) $E_{\mathrm{in}} = 6~\mu$J and $z_{\mathrm{sf}} \approx 100~\mu$m, (c) $E_{\mathrm{in}} = 6~\mu$J and $z_{\mathrm{sf}} \approx 200~\mu$m. For each condition, the sub-figures for line-averaged surface roughness profiles illustrate different regions of the cut surface: (a1, b1, c1) transverse cross-sections showing the modification track morphology, (a2, b2) unmodified surface regions adjacent to the scan path, (b3, c2, c3) laser-scanned regions exhibiting the modification-induced surface texture, and  (a3, c4) residual SiC fragments remaining on the processed tracks.}
\label{fig9}
\end{figure}

In Fig. 9, optical microscopic images describe the modified layer surface and its height profiles with line-averaged surface roughness ($R_\mathrm{a}$) across the beam path region (a1, b1, c1), unmodified region (a2, b2), and the beam passed region (b3, c2, c3), while fragments on the beam path region (a3, c4).
The transverse cross-section profile for all images describes the same peak periods originating from the constrained scanning pitch $W$ = 30 $\mu$m (a1, b1, c1). The lower height regions between the scanned paths correspond to areas that were not irradiated by the laser and therefore remain flat (a2, b2).

When comparing Fig. 9a and Fig. 9b at the same pulse energy, the width of the scanned beam path region increases as the focal depth is set deeper. Especially, effective focal depth with the pulse energy makes a relatively smoother surface profile on the beam path region (a3, b3).
In contrast, a comparison between Fig. 9b and 9c shows that, at a similar modification depth, increasing the pulse energy amplifies height fluctuations in the surface profile, resulting in a larger roughness amplitude under higher pulse energy and deeper focusing conditions.
And the beam paths overlap the scan pitch, where the region between the middle of the scanned paths has a rough surface profile (c2). The surface profile on the beam path region has more roughness than the effective beam pulse energy (b3, c3). The fragments on a beam path region (light gray) have a similar roughness level, while the profile exhibits a lower spatial frequency than the beam path region without fragments (c3, c4).

Fig. 10 provides higher magnification SEM images of the surface regions discussed in Fig. 9b,c, revealing the fine-scale texture responsible for the measured roughness trends. 
Under the $E_{\mathrm{in}} = 3~\mu$J condition, the unmodified region (Fig. 10c) remains featureless, whereas the laser-scanned beam path regions (Fig. 10d,e) exhibit a distinct micro-textured morphology consistent with shallow surface ablation along the track.
In contrast, at $E_{\mathrm{in}} = 6~\mu$J (Fig. 10h-j), the beam path regions show more pronounced surface disruption, including locally intensified roughening in the double-scanned area (Fig. 10h) compared with the single-scanned area (Fig. 10i). 
In the double-scanned region, the surface appears locally stratified compared with the single-scanned area, suggesting that the trace left by the first beam pass modifies the local optical boundary conditions for the second pass. The second pass perturbs the effective multifocal modification height along the z-direction and forms an additional modified band near the focal-affected zone, consistent with the emergence of a secondary track-like feature on the surface. Furthermore, the surface fragments observed in Fig. 10j are detached from the ablated region on the opposite sliced surface generated during separation.

\begin{figure}[h!]
\centering
\includegraphics[scale=0.4]{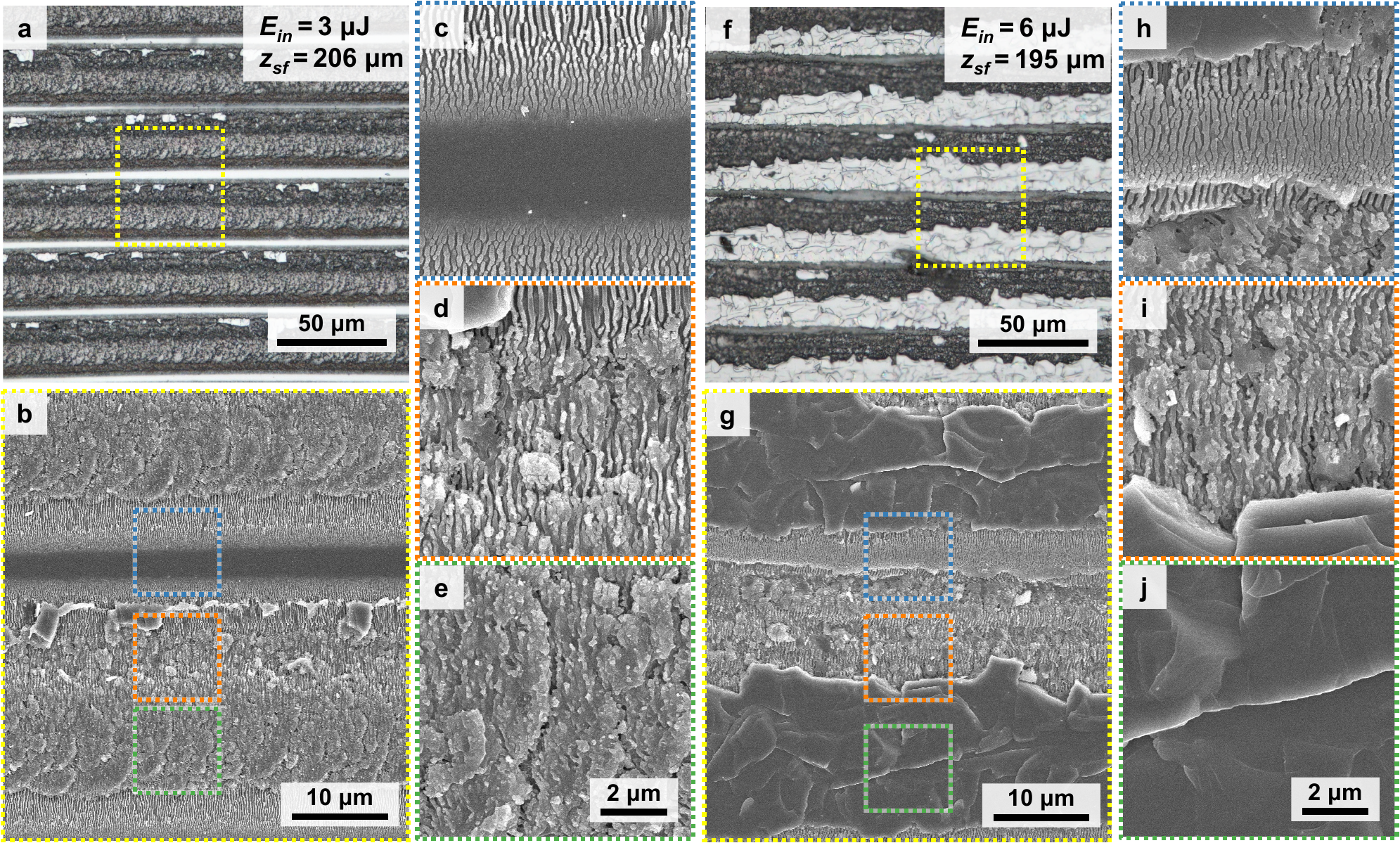}
\caption{(a,f) Optical micrographs of the sliced 4H-SiC wafer surfaces processed at $E_{\mathrm{in}} = 3~\mu$J and 6$~\mu$J, respectively, with comparable self-focusing depths ($z_{\mathrm{sf}} \approx 200~\mu$m) in both cases. (b,g) SEM images taken from the regions indicated in (a) and (f), highlighting the surface morphology between adjacent modification tracks. For the $E_{\mathrm{in}}=3~\mu$J condition, the sub-images of SEM show representative (c) unmodified surface regions and (d,e) laser-scanned regions. For the $E_{\mathrm{in}}=6~\mu$J condition, the sub-images of SEM show (h) double-scanned and (i) single-scanned beam path regions, including (j) residual fragments remaining along the modification tracks.}
\label{fig1}
\end{figure}

\begin{figure}[h!]
\centering
\includegraphics[width=\linewidth]{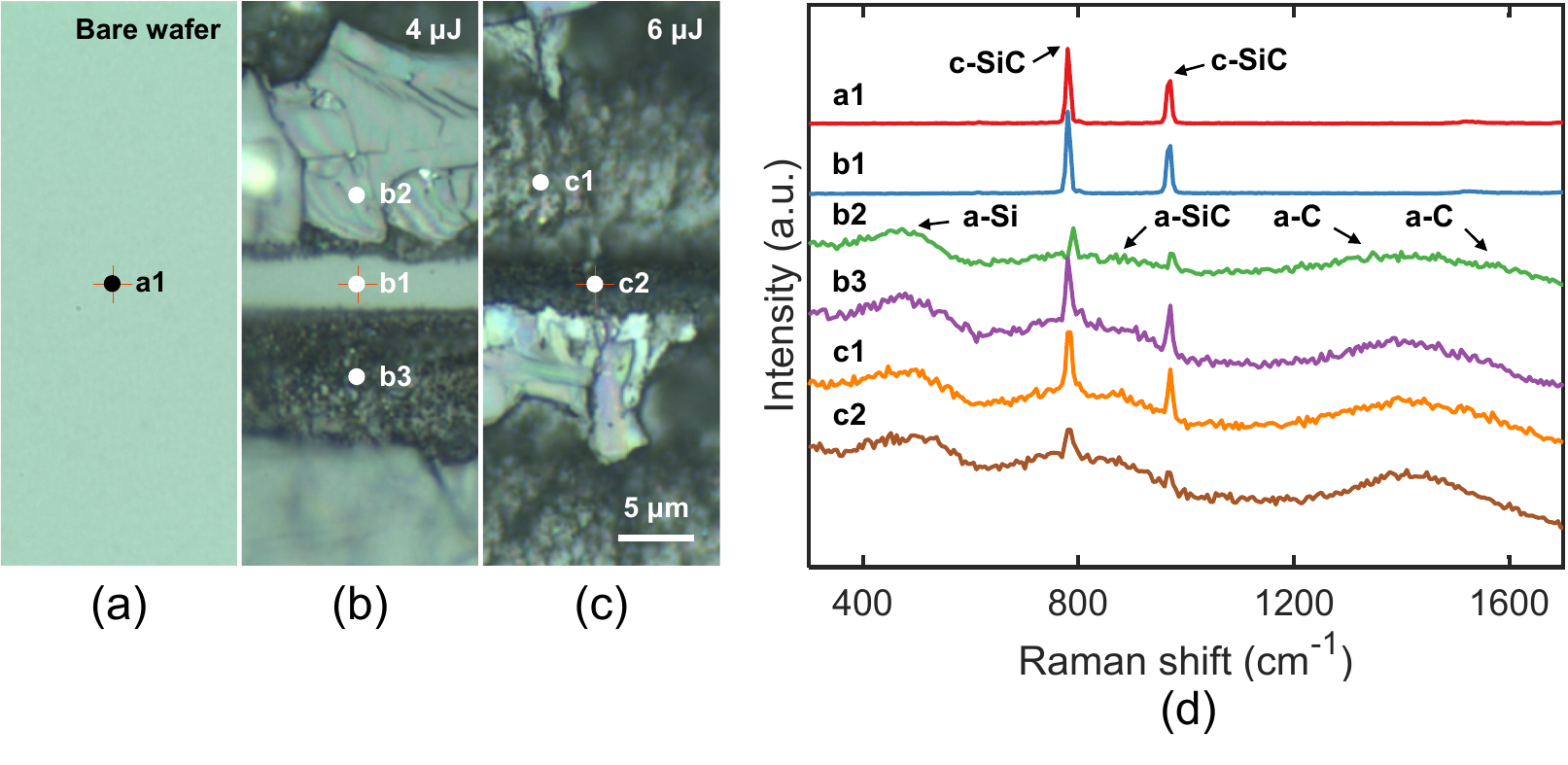}
\caption{Raman spectroscopy analysis of the 4H-SiC crystalline-structure decomposition induced by different pulse energies along the laser beam path. (a) Optical microscopic image of the bare 4H-SiC wafer. 
Modification layer images of samples processed at 
(b) 4 $\mu$J and (c) 6 $\mu$J, showing the Raman measurement locations. 
(d) Corresponding Raman spectra obtained from the marked positions in (a-c).}
\label{fig8}
\end{figure}

\subsection{Microstructural origin of modification width and wafer separability}
\label{3.6}
Raman spectroscopic analysis indicates that the surface fragments originate from the opposite side of the modification layer (Fig. 11), where femtosecond laser irradiation induces 4H-SiC decomposition and microvoid formation. The associated localized thermal expansion generates crack-like features within the decomposed layer, and these partially detached regions are subsequently released from adjacent layers during wafer separation.

The effect of beam modification is compared on the bare wafer and the beam scanning with 4 and 6 $\mu$J pulse energies. The two sharp Raman peaks characteristic of crystalline 4H-SiC (c-SiC) at 781 and 969 cm$^{-1}$ are clearly observed on the bare wafer and in the unmodified regions (a1, b1). In contrast, the laser-modified layer in Fig. 11b,c exhibits a broad and weakened Raman response between these two phonon modes, consistent with the formation of an amorphous SiC phase (a-SiC) $\cite{Naka16, Liu25}$. In addition, the darkened modified regions display a broad band near 480 cm$^{-1}$ indicative of amorphous Si (a-Si), together with D- and G-band features at approximately 1350 and 1580 cm$^{-1}$ corresponding to sp$^{2}$-rich amorphous carbon (a-C), respectively $\cite{Naka16, Du25}$. 
Relatively reduced c-SiC peaks are well explained in the beam path overlap region (c2), while a similar peak distribution appears in the fragment surface on the beam path region (b2). We confirmed that the thickness of the detached fragment was estimated to be < 2 $\mu$m from the surface roughness line profile. 
Considering the depth resolution and optical penetration of Raman spectroscopy, the spectra collected on a surface fragment inevitably include contributions from both the fragment surface and the modified layer beneath it. This reveals that the subsurface region under the fragment is also transformed, indicating that the slicing process produces valley-type morphology not only at the upper portion of the scan track but also along its lower boundary.

\begin{figure}[h!]
\centering
\includegraphics[scale=0.5]{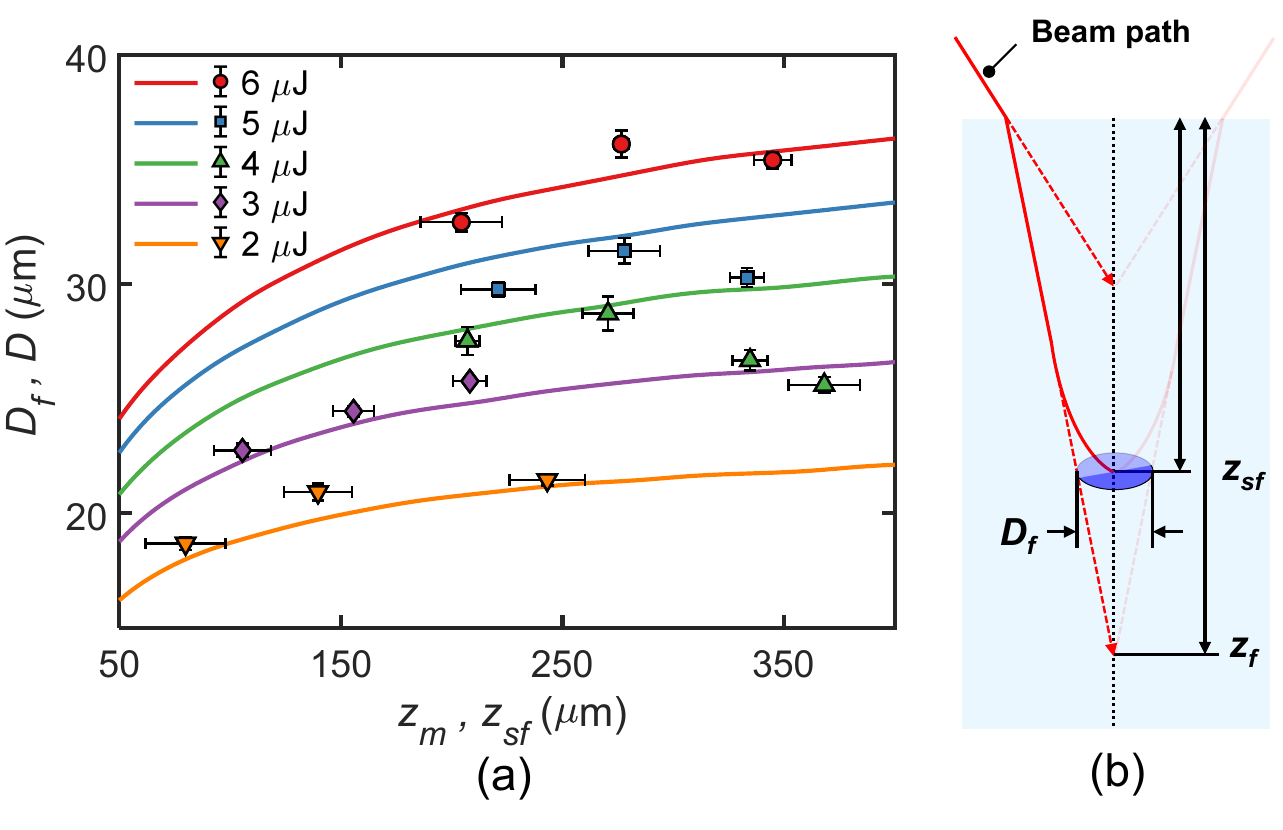}
\caption{(a) Relationship between the modification depth ($z_\mathrm{m}$) and the modification width ($D$). Data points show the mean with error bars denoting $\pm 1$ SD error bars on both axes from the mean value (n=5). Solid lines represent the calculated beam diameter ($D_\mathrm{f}$) at the self-focusing depth ($z_\mathrm{sf}$) based on beam propagation geometry with the modified Marburger equation. (b) The $D\mathrm{_f}$ is estimated at $z_\mathrm{sf}$ by linearly extrapolating the focused conical beam converging toward the geometrical focus ($z_\mathrm{f}$). The $D_\mathrm{f}$ represents the geometrically projected beam width at $z_\mathrm{sf}$, based on the conical profile extending toward $z_\mathrm{f}$.}
\label{fig9}
\end{figure}

Modification width ($D$) on the beam path region follows the analytical model in Fig. 12a. For the pre-collaped diameter ($D_\mathrm{f}$), we assumed that the horizontal modification width scales proportionally, changing by the geometric convergence angle, with the axial offset between the $z_\mathrm{f}$ and the $z_\mathrm{sf}$ in Fig. 12b. The solid lines depict the calculated $D_\mathrm{f}$ in the analytical model by the $z_\mathrm{f}$ with the pulse energy. 
This scaling was valid within the EPZ. However, near and beyond the SFT, the modification width decreased, as observed in the experimental data for $z_\mathrm{m}$ > 320 $\mu$m. 
This reduction reflects a diminished above-threshold radial extent and increased beam energy loss, which can be attributed to spherical aberration-induced elongation of the focal region at larger depths. This elongation redistributes the beam energy along the axial direction. while, for a deeply focused Gaussian beam, partial absorption of sub-threshold influence at the beam periphery further depletes the radial energy content. Together, these effects accelerate the contraction of the effective modification area in the radial direction.

The results were cautiously interpreted as an observed decrement in $D$ at larger self-focusing depths. First, as $z_\mathrm{sf}$ moves deeper, which is closer to $z_\mathrm{f}$ by decreasing the $I_{\mathrm{in}}/I_{\mathrm{cr}}$ as in the modified Marburger formula. In the beam intensity profile across the beam width, the Kerr-driven collapse produces a more core-concentrated on-axis profile as a sharper peak with weakened wings. The radius of the above-threshold region can shrink even while the peak grows. Second, proximity to the exit surface at large depth likely reduces the bulk energy deposition through transmission and partial reflections. Any standing wave modulation near the interface has a sub-micron scale, which tends to modulate the core of the beam profile rather than broaden the effective width, consistent with a net narrowing \cite{Amor14}. These effects give a self-consistent explanation for the width reduction at greater depths, especially near SFT, while the back surface coating could be applied to isolate further the interface contribution \cite{Lenh07}.

\begin{figure}[h!]
\centering
\includegraphics[scale=0.6]{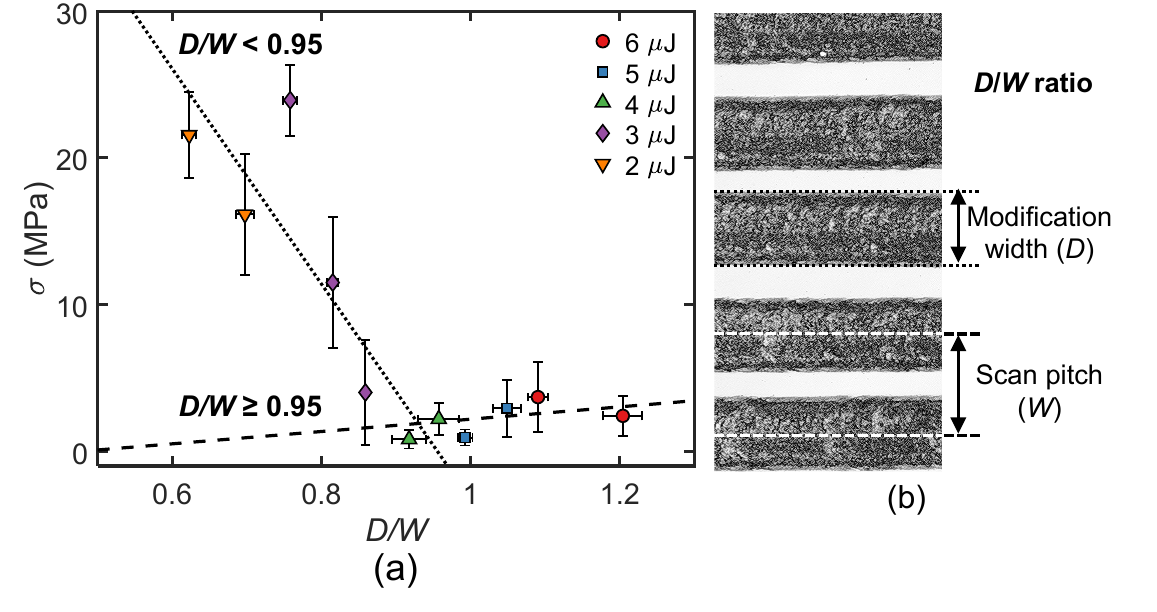}
\caption{(a) Relationship between normalized modification width ($D/W$) and wafer separation stress ($\sigma$). Mean values with $\pm 1$ SD error bars on both axes from the mean value (n=5). All data points represent conditions where separation was performed within the effective processing zone. Linear trends were fitted separately below and above the threshold $D/W$ of 0.95, indicating distinct trends in the correlation. (b) Optical microscopic image illustrating the modification width ($D$) along the laser beam path region, and the scan pitch ($W$) defined as the gap between the center lines of adjacent beam paths.}
\label{fig10}
\end{figure}

Within the EPZ, the modification width directly affects wafer separability. Fig. 13 shows the relation between the separation stress and the normalized modification width $D/W$. At lower pulse energy, the irradiance can exceed the MOT only if the beam contracts more tightly at the collapse point, when the effective beam radius is smaller. Consequently, the wafer separation requires a lower separation stress for the wider modification width as close to 0.95 of the $D/W$. Then, the separation stress becomes constant below 5 MPa within the $D/W$ $\geq$ 0.95 range. Overlapping of the beam path region does not lead to re-bonding between adjacent interfaces within the modification layer, as evidenced by the consistently low separation stress observed even when modification tracks overlap. As seen in Raman spectroscopic results, multiple passages of the beams rather decomposed the crystal structure, reducing the crystalline peaks, consistent with the spectroscopic results. As the irradiance increases toward the optimal range for slicing, the surface texture parameters also reflect this microstructural evolution. 
The $S_\mathrm{ku}$ and the magnitude of the negative $S_\mathrm{sk}$ increase, indicating a growing population of narrow, valley-type depressions along the sliced surface. The trend aligns with the reduction of effective modification width ($D$) relative to the scan pitch ($W$), yielding an increase in the $D/W$ ratio. When considering the overall processability of femtosecond laser slicing, a beam-irradiance condition near $D/W \approx 0.95$ provides an optimal balance. At this ratio, the modification zone spans a sufficient depth to ensure continuous slicing while the valley-type surface features remain fine and uniformly distributed, thereby minimizing both surface damage and the required separation stress.

\begin{figure}[h!]
\centering
\includegraphics[scale=0.6]{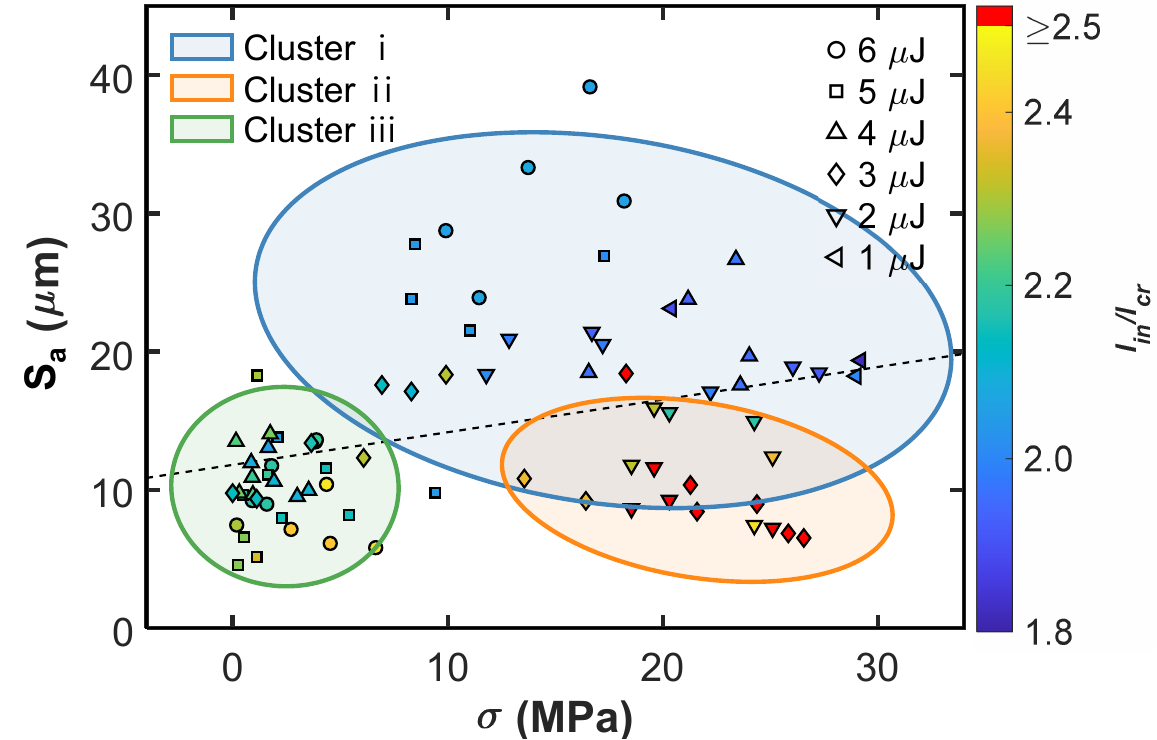}
\caption{Correlation between the separation stress ($\sigma$) and the areal-average surface roughness ($S_\mathrm{a}$) with three clusters overlaid on all separable samples. The color of each symbol represents the normalized irradiance ($I_{\mathrm{in}}/I_{\mathrm{cr}}$). Solid curves denote 95\% confidence ellipses that summarize the spread of each group.}
\label{fig11}
\end{figure}

\begin{figure}[h!]
\centering
\includegraphics[width=\linewidth]{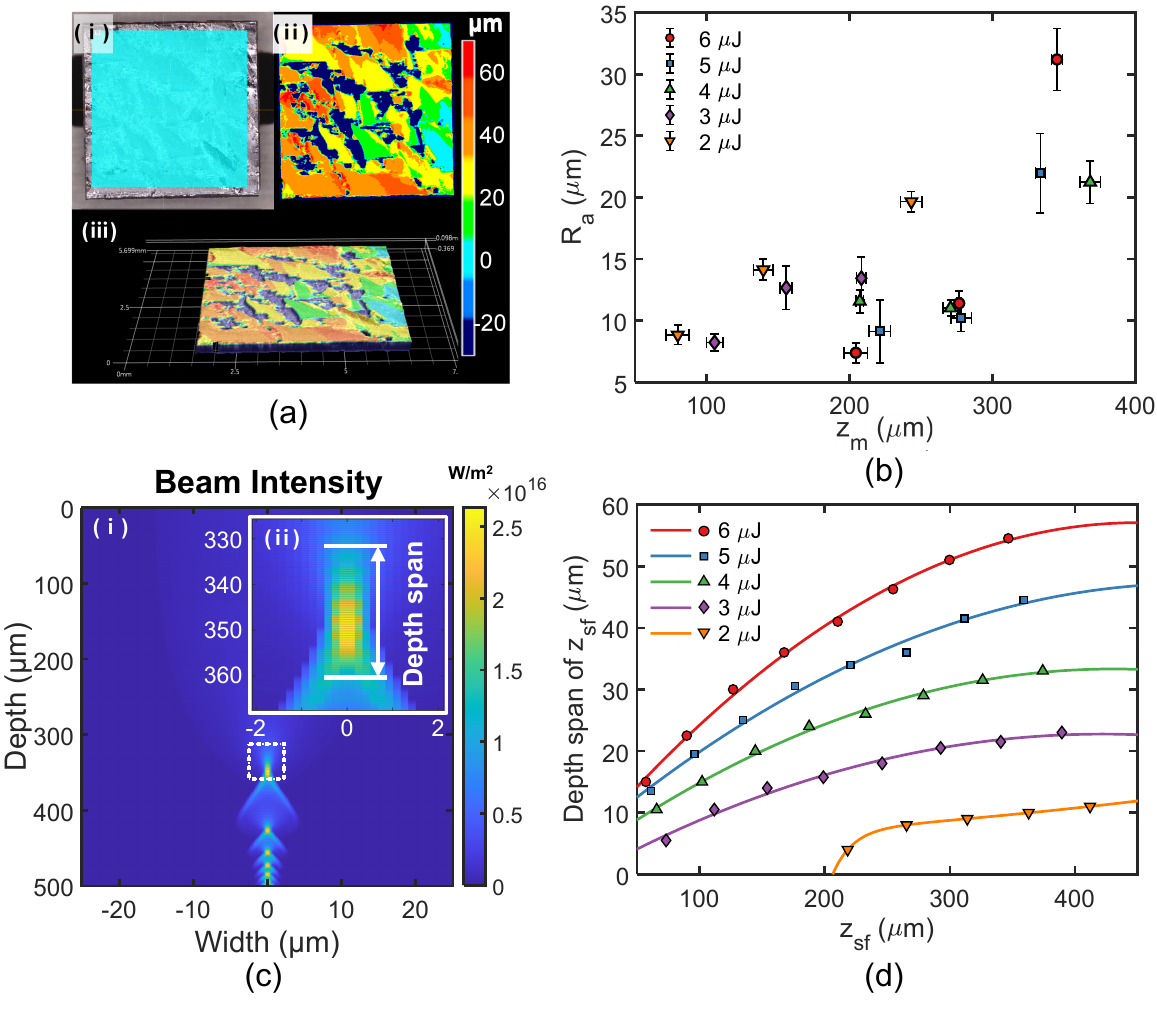}
\caption{(a) Surface roughness at the region of interest of a separated surface (i), and its 2D and 3D surface profiles (ii, iii). (b) Areal-averaged surface roughness at the modification depth by the magnitude of pulse energy. Mean values with $\pm 1$ SD error bars on both axes from the mean value (n=5). (c) Depth span at the self-focusing depth in simulation over $I_\mathrm{cr}$, and (d) depth spans by the self-focusing depth and the magnitude of the pulse energy.}
\label{fig12}
\end{figure}

\subsection{Correlation between self-focusing depth, surface roughness, and separation stress}
\label{3.7}
Another process quality evaluation parameter, $S_\mathrm{a}$, distinguishes the processability together with the $\sigma$. The three clusters map to separable processing windows on the $\sigma$-$S_\mathrm{a}$ plane in Fig. 14. The high $\sigma$-high $S_\mathrm{a}$ group for the cluster i (blue), the high $\sigma$-low $S_\mathrm{a}$ group for the cluster ii (orange), and the the low $\sigma$-low $S_\mathrm{a}$ group for the cluster iii (green). The cluster groups and color coding are consistent with those used in Fig. 8 and Table 1.

All samples were grouped into three clusters using the k-medoids on the standardized parameters of $\sigma$, and $S_\mathrm{a}$. Medoids are depicted as cluster centers, and the spread of each cluster is summarized by 95\% confidence ellipses from the covariance for the sparse groups. The choice $k$=3 of the number of clusters is supported by internal validity. The mean silhouette over $k$ $\in$ \{2,3,4,5,6\} peaked at $k$=3, with scores 0.621, 0.671, 0.652, 0.665, and 0.546 for $k$ = 2,3,4,5, and 6, respectively. The incremental changes beyond $k$=3 are modest and mainly reflected subdivisions within the $\sigma$-$S_\mathrm{a}$ regions rather than new physically distinct clusters. Accordingly, the $k$=3 clustering is consistent with the zone classification in the processability map. 

The normalized irradiance is shown in the color of symbols only and does not affect the labels. The blue cluster i shows a wide range of separation stress with the high surface roughness. The color of the symbols is close to blue, $I_{\mathrm{in}}/I_{\mathrm{cr}}$ $\leq$ 2.1. The irradiance reached at the modification depth is not sufficient to keep the slicing quality regardless of the incident powers, both 1 and 6 $\mu$J conditions. The low magnification for the precise slicing is related to the increment of the separation stress for the samples within the zone i of the TDZ, Near-IPZ. 
In the same way for the orange cluster ii, the colors of the symbols are yellow and red,  $I_{\mathrm{in}}/I_{\mathrm{cr}}$ $\geq$ 2.3. The symbols indicated the power conditions with 2 and 3 $\mu$J at zone ii of the near-SDZ. High irradiance of the self-focusing depth near the surface is better to minimize the surface roughness of the separation layer, but it requires diminishing scan pitch to minimize the separation stress. The green cluster iii is located in the region of low separation stress with low surface roughness, where the color of the symbols indicates within the  2.1 $\leq$ $I_{\mathrm{in}}/I_{\mathrm{cr}}$ $\leq$ 2.3.

At the macroscopic level, the behavior of the areal-averaged surface roughness ultimately traces back to the focusing depth. In Fig. 15a, the $S_{\mathrm{a}}$ is calculated within the sky blue region of interest for all samples, and representative 3D profiles are illustrated with the texture. For a fixed pulse energy, $S_{\mathrm{a}}$ increases as the modification depth is set deeper, in parallel with the rise of the on-axis peak irradiance at self-focusing depth in Fig. 15b. This trend is consistent with cumulative Kerr nonlinearity, which produces a larger axial extent of vertical modification along the beam path region, together with imprinting rougher features on the separated surface. In Fig. 15c, above-threshold depth span is defined $L_{\mathrm{th}}$ at the first self-focusing point as the axial extent for which $I_{\mathrm{z}}$ $\geq$ $I_{\mathrm{cr}}$. The $L_{\mathrm{th}}$ is described in Fig. 15d, which can exceed 50 $\mu$m under the operating conditions. The long above-threshold depth spans imply a substantial modified volume, which can lead to significant kerf loss during subsequent planarization to achieve a flat surface.

\section{Conclusion}
\label{sec4}
We investigated the femtosecond laser slicing and layer separation within the processability map. The processability map describes the effective processing zone within the physically defined borders of the surface damage threshold (SDT), self-focusing threshold (SFT), and modification onset threshold (MOT). The classified zones have been linked with the quality parameters for the separation stress ($\sigma$) and the areal-averaged surface roughness ($S_\mathrm{a}$). The processability map based on the experiments was developed and validated with analytical and simulation models as follows.

\begin{itemize}
\item Approximation of the Marburger formula to the focused beam was performed to replace the parameters of the collimated beam from the original equation. In the modified equation, the critical irradiance ($I_\mathrm{cr}$) is assumed to be a constant for both the surface ablation and plasma generation within a medium. 
\item Validation of the self-focusing depth ($z_\mathrm{sf}$) was performed on the experimental, analytical, and simulation models. Then, experimental results are plotted on the processability map of the incident pulse energy ($E_\mathrm{in}$) versus the $z_\mathrm{sf}$, underlaid on the normalized irradiance ($I_\mathrm{in}/I_\mathrm{cr}$) background. The threshold boundaries separate the results into three categories by
geometric and optical limits.
\item Microscopic characteristics (within the beam path scale) show the decomposition of the crystalline structure by Raman spectroscopic results. Separation stress is evaluated to screen the optimal slicing conditions by the normalized modification width within the microscopic scale. 
\item Macroscopic characteristics (within the modification layer scale) in data clustering are depicted on the quality parameters $\sigma$-$S_\mathrm{a}$ map, which matched the processability map. In addition, the $S_\mathrm{a}$ follows the above-threshold depth span ($L_\mathrm{th}$). Initial crack location follows within the depth span region.
\end{itemize}

Establishing a physics-based correlation between laser inputs and slicing outcomes is essential for process optimization. The clearer the mapping from inputs to intermediate, physically meaningful descriptors, e.g.,$E_\mathrm{in}$, $z_{\mathrm{sf}}$, and ${I_\mathrm{in}/I_{\mathrm{cr}}}$, and finally to macroscopic quality metrics for the $\sigma$, and $R_{\mathrm{a}}$, the more effectively can regulate parameters for processability. This is especially critical in femtosecond machining, where strong nonlinearity obscures quantifying robust indicators that capture the Kerr-driven collapse. 
Furthermore, a data library with the quantified markers amplifies the usefulness of data-driven tools using AI models in uncovering latent patterns within nonlinear regimes and accelerating recipe search. Practically, mid-process measurements, such as wafer separation stress and monitoring of surface roughness of the separated layer, can be integrated into a closed loop to evaluate the wafer slicing process in real-time, enabling adaptive updates and improving cost and yield.

\section*{Acknowledgment}
This research was supported by “Development of fine pitch micro-bump bonding process and equipment with high-efficiency for high performance semiconductor package” through the National Research Foundation of Korea (NRF) grant funded by the Korea government (MSIT) (RS-2024-00431837) and a grant of the Basic Research Program funded by the Korea Institute of Machinery and Materials (grant number: NK254A, Project Title: Development of Core Technologies for Advanced Chiplet Packaging Equipment).

% \section*{CRediT author statement}

% \textbf{Dong Hee Kang:} Conceptualization, Formal analysis, Investigation, Methodology, Visualization, Writing – original draft \textbf{Jaeseung Lim:} Data curation, Investigation, Methodology, Writing – review & editing \textbf{Mishfaqur Rahman:} Formal analysis, Investigation, Methodology, Validation, Visualization, Writing – original draft \textbf{Seongheum Han:} Funding acquisition, Project administration, Resources, Writing – review & editing \textbf{Jae-Hak Lee:} Funding acquisition, Investigation, Project administration, Supervision, Writing – review & editing \textbf{Seungman Kim:} Conceptualization, Formal analysis, Funding acquisition, Investigation, Methodology, Project administration, Supervision, Writing – review & editing \textbf{Jihoon Jeong:} Conceptualization, Funding acquisition, Investigation, Resources, Supervision, Writing – original draft, Writing – review & editing

{\newpage}
\appendix
\section{Kerr-induced self-focusing in the femtosecond laser}  
\label{app1}
In ultra-short pulsed laser beams, the optical path should be considered for Kerr effects for accurate modification depth at the self-focusing point. The self-focusing effect is that the beam modifies the refractive index and concentrates itself inside the material. The refractive index considering Kerr effects is described \cite{Boyd08}, 
\begin{equation}
n = n{_0} + n_\mathrm{Kerr}I.\\    
\end{equation}
The total refractive index ($n$) includes the linear ($n_0$) and nonlinear ($n_\mathrm{Kerr}$) refractive indices and the laser pulse intensity ($I$). The critical power for the self-focusing ($P_\mathrm{cr}$) is given by \cite{Boyd08},
\begin{equation}
P{_\mathrm{cr}} = \frac{3.77\, \lambda{_0^2}}{8 \pi n_0 n_\mathrm{Kerr}},
\end{equation}
defines the threshold above which the Kerr effect induces beam collapse in a nonlinear medium.
The critical power of the Kerr effect is 310 kW, when the $n_\mathrm{Kerr}$ is the  3.72 × 10$^{-19}$ m$^2$/W of the 4H-SiC around the 1,000 nm of beam wavelength \cite{Guo21}. The peak power in the setup is compared to the equation below, 
\begin{equation}
P{_\mathrm{peak}} = \frac{P_{\mathrm{in}}}{f \cdot \tau{_l}}
\end{equation}
The peak power of a single pulse of the femtosecond laser is 12.5 MW at 1 W of the incident power ($P_\mathrm{in}$) condition, the light intensity exceeds around 40 times the critical power for the Kerr effect. Therefore, the nonlinear optical phenomenon is inevitable in laser slicing using the femtosecond laser to control the precision depth control.

\section{Feasibility of the modified Marburger formula for self-focusing effect approximation}
\label{app2}
The relation of the self-focusing depth with the beam characteristics is described by the semi-empirical Marburger equation,   
\begin{equation}
z_{\mathrm{sf}} = \frac{0.367\, k_0 w_{0}^{2}}{\sqrt{\left[ \sqrt{P_{\mathrm{in}} \left/ P_{\mathrm{cr}} \right.} - 0.852 \right]^2 - 0.0219}}.
\end{equation}
Here, $k_0$ is the vacuum wavenumber, $k_0=2\pi /\lambda_0$. The Marburger equation assumes a collimated beam and uses the diffraction length (Rayleigh range), $z_\mathrm{R}=\pi w_{0}^{2}/\lambda_0$, as the characteristic axial scale in the numerator. To justify the substitution of the Rayleigh range $z_\mathrm{R}$ with the linear focal distance $z_\mathrm{f}$ in the self-focusing distance formula, we consider the spatial rate of change of beam radius in both collimated and focused beam configurations.
For a collimated Gaussian beam in a homogeneous medium of refractive index $n$, the characteristic convergence half-angle is given by
\begin{equation}
\theta = \frac{w_0}{z_\mathrm{R}}.
\end{equation}
Far from the waist of a collimated Gaussian beam ($|z|\gg z_R$), the radius changes approximately linearly with z as \(w(z)\approx \theta |z|\). Hence, the asymptotic slope is
\begin{equation}
\left. \frac{d w}{d z} \right|_{|z| \gg z_\mathrm{R}} \simeq \theta.
\end{equation}
In the focused beam configuration, to avoid the self-referential definition with nonlinear terms switched off, pre-collapse radius (linear beam radius) $w_\mathrm{lin}(z)$ is evaluated from a purely linear propagation model. It represents the virtual radius the beam would have at axial position $z$ in the absence of nonlinear interaction. We evaluate the pre-collapse radius using the geometric convergence angle from the geometric depth,
\begin{equation}
w_\mathrm{lin}(z) \simeq \sqrt{w_0^2 + (\theta \cdot |z-z_{0}| )^2}.
\end{equation}
For small $\theta$, $\tan\theta \simeq \theta = w_0/z_R$, which recovers the Gaussian beam radius $w(z)=w_0\sqrt{1+((z-z_f)/z_R)^2}$, when the waist is at $z=z_f$. We use the linear radius $w_\mathrm{lin}$ evaluated at $z=z_{\mathrm{sf}}$ to compute the normalized irradiance ($I_{\mathrm{in}}/I_{\mathrm{cr}}$) in the focused beam extension of Marburger formula. Because both $z_\mathrm{R}$ and $z_\mathrm{f}$ set the axial scale over which the beam radius varies significantly under linear optics, it is physically reasonable in the high power limit $P_{\mathrm{in}}\gg P_{\mathrm{cr}}$ to replace $z_\mathrm{R}$ with $z_\mathrm{f}$ in the self-focusing relation.

In the original Marburger equation, the onset of nonlinear beam collapse is parameterized by the power ratio, $P_\mathrm{in}/P_\mathrm{cr}$. This expression is appropriate for a collimated beam, where the entire beam contributes to nonlinear self-focusing. However, the spatially varying beam profile leads to highly localized intensities in tightly focused beams, such as the femtosecond laser for slicing. In the case of a tightly focused beam, the incident irradiance $I_\mathrm{in}$ becomes more relevant than the total incident power $P_\mathrm{in}$, as the irradiance increases inversely with the beam cross-sectional area along the propagation path. Since plasma generation is directly driven by the local intensity at the focal region, we adopted the irradiance ratio $I_\mathrm{in}/I_\mathrm{cr}$ in place of the original power ratio $P_\mathrm{in}/P_\mathrm{cr}$ used in the Marburger equation.

Although the numerical coefficients of 0.367, 0.852, and 0.0219 in the Marburger equation were originally obtained for a collimated Gaussian beam, we retain these constants in the focused beam extension for the present configuration. For the NA=0.42 optics in 4H-SiC, the modified relation accurately reproduces the self-focusing depth when compared against ray optics simulations and experimental results at $E_{\mathrm{in}}$ = 2 to 6 $\mu$J, with agreement within the experimental uncertainty. This empirical validation justifies using the original coefficients without additional calibration in the operating range. The resulting processability map supports these assumptions. 

In the analytic model, collapse requires a normalized irradiance exceeding a material-dependent critical level at least ${I_\mathrm{in}/I_{\mathrm{cr}}}\geq 1.54$. Whereas in experiments, separation was observed for ${I_\mathrm{in}/I_{\mathrm{cr}}}\geq 1.9$. The upward shift of the experimental threshold is consistent with plasma dispersion, free carrier absorption, and scattering. 

The applicability of the analytical model is restricted to the MOT-SFT-SDT window: the lower irradiance bound is set by MOT, the upper irradiance bound by SDT, and the optical-path limit for modification by SFT, beyond which surface ablation or cracking occurs. Within this map, the normalized collapse distance follows the inverse square root scaling with relative irradiance,
\begin{equation}
z_{\mathrm{sf}}/z_{\mathrm{f}} \propto \frac{1}{\sqrt{I_{\mathrm{in}} / I_{\mathrm{cr}}}},
\end{equation}
as expected from the original Marburger equation. The processability map would require recalibration of the constants in the cases, e.g., extreme NA, spherical aberration, a deep focusing with strong aberration, or ${I_\mathrm{in}/I_{\mathrm{cr}}}$ well above SDT, and plasma-induced defocusing.

\section{Ray optics simulation and methodology}
\label{app3}
We model femtosecond-pulse self-focusing in bulk 4H-SiC by coupling a Kerr-induced refractive index field to a geometric-optics ray tracer and then reconstructing the spatiotemporal intensity from the ensemble of rays; nonlinearity enters through the instantaneous Kerr term.

\subsection{Optical field and Kerr index}
\label{app3.1}

The refractive index is taken as
\begin{equation}
n(x,y) = n_{0} + n_{Kerr}\, I(x,y)
\label{eq:Kerr}
\end{equation}
with \(n_0=2.55\) and \(n_\mathrm{Kerr}=3.72\times10^{-19}\ \mathrm{m^2/W}\) for 4H-SiC at \(1040\,\mathrm{nm}\). The incident pulse has energy \(E_{\mathrm{pulse}}=\) 2 to 6 $\mu$J and duration \(\tau=400~\mathrm{fs}\). Because \(R=0.19\) in the simulations reported here, the transmitted energy is \(E_{\mathrm{trans}}=(1-R)E_{\mathrm{pulse}}\). A paraxial Gaussian beam is prescribed, focused to the nominal axial position \(y=z_f\) with waist diameter \(w_0^{(\mathrm{diam})}=2.746~\mu\mathrm{m}\) (waist radius \(w_0=w_0^{(\mathrm{diam})}/2\)). Inside the medium the wavelength is \(\lambda_m=\lambda_0/n_0\) with \(\lambda_0=1040~\mathrm{nm}\). The depth-dependent beam radius is
\begin{equation}
w(y) = w_0\,\sqrt{1 + \left(\frac{y - z_f}{z_R}\right)^2}\,,
\end{equation}
\begin{equation}
z_R = \frac{\pi w_0^2}{\lambda_m}
    = \frac{\pi n_0 w_0^2}{\lambda_0}\,.
\end{equation}
This intensity used to seed the nonlinear index (i.e., the intensity field that defines n(x,y) before ray tracing) is the standard Gaussian profile with the correct depth dependence,
\begin{equation}
I_{\text{seed}}(x,y)
= \frac{E_{\text{trans}}}{\tau\,\pi\!\left(\frac{w_0}{2}\right)^2}
  \left(\frac{w_0}{w(y)}\right)^{\!2}
  \exp\!\left[-\frac{2x^2}{w(y)^2}\right].
\end{equation}
This field is evaluated on a two-dimensional Cartesian grid
\((x,y)\in[-25,25]~\mu\mathrm{m}\times[0,500]~\mu\mathrm{m}\)
with \(N_x=300\) and \(N_y=2000\) nodes, yielding uniform spacings
\(\Delta x = \frac{50~\mu\mathrm{m}}{N_x-1}\) and
\(\Delta y = \frac{500~\mu\mathrm{m}}{N_y-1}\).
The nonlinear refractive index is modeled as
\begin{equation}
  n(x,y) = n_0 + n_\mathrm{Kerr}\, I_{\text{seed}}(x,y),
\end{equation}
and its spatial gradients
\(\nabla n = \bigl(\partial n/\partial x,\; \partial n/\partial y\bigr)\)
are computed numerically on the grid and interpolated to the off-grid ray
positions using MATLAB's \texttt{griddedInterpolant}.

\subsection{Ray-based propagation}
\label{app3.2}

Beam propagation is modeled in geometric optics via the eikonal (ray) system derived from Fermat’s principle in an inhomogeneous medium:
\begin{equation}
  \frac{d\mathbf{r}}{ds}=\mathbf{v}, 
  \qquad 
  \frac{d\mathbf{v}}{ds}=\frac{1}{n\!\left(\mathbf{r}\right)}\,\nabla n\!\left(\mathbf{r}\right),
  \label{eq:eikonal}
\end{equation}
with the unit-speed constraint \(\lVert \mathbf{v}\rVert=1\) enforced at every step. We launch \(N_r=500\) rays at the entrance plane \(y=0\), with initial lateral positions \(x_0\in[-25,25]~\mu\mathrm{m}\) uniformly spaced (the central ray is omitted to avoid symmetry-locking artifacts), and initial direction \(\mathbf{v}_0=(0,1)\). Numerical integration uses an explicit forward-Euler update with fixed arclength increment \(ds=0.5~\mu\mathrm{m}\) for 1200 steps; after each update, \(\mathbf{v}\) is renormalized to satisfy \(\lVert \mathbf{v}\rVert=1\). Rays that exit the computational domain are terminated.

\subsection{Energy-conserving intensity reconstruction}
\label{app3.3}

The spatiotemporal intensity is reconstructed by superposing the energy carried by each geometrical ray along its path. Each ray \(j\) is assigned a launch weight \(w_j\) proportional to a Gaussian in its initial lateral coordinate \(x_{0,j}\),
\begin{equation}
  \tilde w_j \;\propto\; \exp\!\left(-\frac{x_{0,j}^2}{2\sigma_w^2}\right),
  \qquad \sigma_w = 10~\mu\mathrm{m},
\end{equation}
and we normalize the weights so that
\begin{equation}
  w_j \;=\; \frac{\tilde w_j}{\sum_{\ell}\tilde w_\ell},
  \qquad
  \sum_{j} w_j = 1.
\end{equation}
The energy attributed to ray \(j\) is then
\begin{equation}
  E_j \;=\; w_j\,E_{\mathrm{trans}}.
\end{equation}
Along the discrete trajectory of ray \(j\), indexed by the set of in-bounds arclength samples \(k\in \mathcal{S}_j\) at positions \((x_k,y_k)\), we distribute the ray energy uniformly (no absorption in this run). Writing \(N_j = |\mathcal{S}_j|\), the elemental energy per sample is \(E_j/N_j\). At each sample we deposit this energy into the Eulerian \((x,y)\) grid by a compact, normalized \(5\times 5\) Gaussian kernel \(K\) (unit sum) centered at the nearest grid node. Dividing by the voxel 4-volume \(\Delta x\,\Delta y\,\tau\) converts deposited energy to intensity. Concretely, the intensity increment at grid node \((x_i,y_m)\) due to sample \((j,k)\) is
\begin{equation}
  \Delta I_{i,m}^{(j,k)}
  \;=\;
  \frac{E_j/N_j}{\tau\,\Delta x\,\Delta y}\;
  K\!\bigl(x_i - x_k,\; y_m - y_k\bigr),
\end{equation}
and the total intensity field is the superposition
\begin{equation}
  I(x_i,y_m) \;=\; \sum_{j}\;\sum_{k\in\mathcal{S}_j} \Delta I_{i,m}^{(j,k)}.
\end{equation}
By construction, the deposition is energy-conserving on the mesh:
\begin{equation}
  \sum_{i,m} I(x_i,y_m)\,\Delta x\,\Delta y\,\tau
  \;=\; E_{\mathrm{trans}},
\end{equation}
to within machine precision, which we report as an energy check. The kernel-based deposition both preserves energy and anti-aliases sub-grid ray wandering, yielding smooth, physically interpretable intensity maps without numerical noise.

\subsection{Extraction of self-focusing observables}
\label{app3.4}

From the reconstructed intensity map \(I(x,y)\) we determine the first self-focusing depth \(z_{\mathrm{sf}}\) as the axial position of the global maximum along the optical axis (centerline \(x=0\)):
\begin{equation}
  z_{\mathrm{sf}}
  \;=\;
  \operatorname*{arg\,max}_{y}\, I(0,y).
\end{equation}
The lateral spot size at self-focus is measured by the full width at half maximum (FWHM) of the transverse profile at \(y=z_{\mathrm{sf}}\). Defining
\begin{equation}
  I_{\perp}(x) \;=\; I\bigl(x, z_{\mathrm{sf}}\bigr),
  \qquad
  I_{\max} \;=\; I\bigl(0, z_{\mathrm{sf}}\bigr),
\end{equation}
the FWHM is the distance between the two half-maximum intersections, found by linear interpolation on the discrete grid:
\begin{equation}
  \text{FWHM}
  \;=\;
  x_2 - x_1
  \quad \text{where} \quad
  I_{\perp}(x_1) \;=\; I_{\perp}(x_2) \;=\; \tfrac{1}{2} I_{\max}.
\end{equation}
To quantify the portion of pulse energy participating directly in the self-focusing hotspot, we also report an integrated-energy metric within the “affected region,” defined as the super-level set
\begin{equation}
  \mathcal{A}
  \;=\;
  \bigl\{ (x,y)\;:\; I(x,y) \ge 0.05\, I_{\max} \bigr\}.
\end{equation}
On the discrete mesh \((x_i,y_m)\) with spacings \(\Delta x,\Delta y\) and pulse duration \(\tau\), the energy contained in this region and its fraction of the transmitted pulse are
\begin{equation}
  E_{\mathcal{A}}
  \;=\;
  \sum_{(i,m)\in \mathcal{A}} I(x_i,y_m)\,\Delta x\,\Delta y\,\tau,
  \qquad
  \eta_{\mathcal{A}}
  \;=\;
  \frac{E_{\mathcal{A}}}{E_{\mathrm{trans}}}.
\end{equation}

\subsection{Numerical settings and domains}
\label{app3.5}

All fields are computed on the rectangular domain \([{-}25,25]~\mu\mathrm{m} \times [0,500]~\mu\mathrm{m}\) with a uniform Cartesian discretization of \(N_x = 300\) and \(N_y = 2000\) nodes. The geometric focus is prescribed at \(z_f = 300~\mu\mathrm{m}\). The Rayleigh range employs the waist diameter specified above, and spatial gradients of the nonlinear index \(n(x,y)\) are obtained by centered finite differences on the mesh and evaluated at off-grid ray locations by bilinear interpolation. The ray marching step is \(ds = 0.5~\mu\mathrm{m}\), which resolves the index-gradient curvature while maintaining numerical stability; halving \(ds\) yields indistinguishable values for all reported metrics.

\subsection{Outputs used in this work}
\label{app3.6}

For each simulation we save the reconstructed intensity map \(I(x,y)\), the Kerr-seeded Gaussian map \(I_{\mathrm{seed}}(x,y)\), the nonlinear index \(n(x,y)\), and an energy-conservation diagnostic
\begin{equation}
E_{\mathrm{check}}
= \sum_{i,m} I(x_i,y_m)\,\Delta x\,\Delta y\,\tau .
\end{equation}
We also record the first self-focusing depth \(z_{\mathrm{sf}}\), the corresponding peak intensity \(I_{\max}\), the lateral full width at half maximum (FWHM) evaluated at \(y=z_{\mathrm{sf}}\), and the energy contained in the affected region. All values quoted in the Results are computed directly from these fields produced by the algorithm described above.

\section{Effect of spherical aberration-induced focal elongation}
\label{app4}
The potential influence of spherical aberration caused by the refractive index mismatch at the air/4H-SiC interface was quantitatively evaluated to assess whether it can account for the experimentally observed upstream shift of the modification depth.

For a high refractive index medium, the longitudinal focal shift and elongation can be estimated using geometrical optics relations \cite{Sun05}. The corresponding longitudinal aberration length ($d_{\mathrm{LA}}$) can be expressed as
\begin{equation}
d_{\mathrm{LA}} = \frac{z}{n_{0}}
\left(
\sqrt{\frac{n_{0}^2 - \mathrm{NA}_{\mathrm{eff}}^2}{1 - \mathrm{NA}_{\mathrm{eff}}^2}}
- n_{0}
\right),
\end{equation}
where $n_0$ is the refractive index of 4H-SiC and $\mathrm{NA}_{\mathrm{eff}}$ is the effective numerical aperture determined by the incident beam diameter.

Although the Mitutoyo objective lens has a nominal NA of 0.42, the entrance pupil diameter is 3.36 mm, whereas the incident beam diameter in our experiments was limited to 2 mm. Consequently, the effective numerical aperture is reduced to $\mathrm{NA}_{\mathrm{eff}}$ = 0.25. Using $n_{0} = 2.55$ and a focal position of $z$ = 500 $\mu$m below the 4H-SiC surface, the calculated longitudinal aberration length is $d_{\mathrm{LA}}$ = 13.9 $\mu$m.

The result confirms that spherical aberration-induced focal elongation remains on the order of only a few micrometers under the present experimental conditions. Such a small elongation cannot account for the experimentally observed modification depth shifts on the order of hundreds of micrometers. Therefore, while spherical aberration contributes a static background broadening of the focal region, it cannot be the dominant mechanism governing the systematic upstream shift of the effective modification depth observed in this study.

\section{Notation and definitions for modification depth analysis}
\label{app5}
At a single geometrical focus position on a modification track, suppose that $K$ principal modification points are identified along the depth direction. $K$ typically ranges from 2 to 5, depending on the pulse-overlap condition. We denote their depths by $z_\mathrm{m,k}$ ($k=1,\ldots,K$). The centroid depth of the principal modification distribution within this single site is defined as
\begin{equation}
z_{\mathrm{m},p}=\frac{1}{K}\sum_{k=1}^{K} z_{\mathrm{m},k},
\qquad
s_{\mathrm{m},p}=\mathrm{SD}\left(z_{\mathrm{m},1},\ldots,z_{\mathrm{m},K}\right),
\end{equation}
where $s_{m,p}$ is the standard deviation (SD) of the $K$ principal points at the modification region. $z_{m,p}$ is written without an overbar, which represents a within a track distribution.

For repeated measurements across $n$ tracks, we obtain a set of representative depths $\{z_{m,p}^{(i)}\}_{i=1}^{n}$. The ensemble mean and its dispersion are defined as
\begin{equation}
\bar{z}_{\mathrm{m},p}=\frac{1}{n}\sum_{i=1}^{n} z_{\mathrm{m},p}^{(i)},
\qquad
s_{\bar{z}_{\mathrm{m},p}}=\mathrm{SD}\left(z_{\mathrm{m},p}^{(1)},\ldots,z_{\mathrm{m},p}^{(n)}\right),
\end{equation}
where $\bar{z}_{\mathrm{m},p}$ denotes the ensemble-averaged representative depth and $s_{\bar{z}_{\mathrm{m},p}}$ is the SD across the $n$ measurements.

Among the $K$ principal modification points identified at a single processing site, we define the primary modification depth, $z_\mathrm{m}$, as the incident side first modification point, $z_\mathrm{m} \equiv z_\mathrm{m,1}$, where the indexing is chosen such that $z_\mathrm{m,1}$ corresponds to the closest to the incident surface.
For repeated cross-sectional tracks ($i=1,\ldots,n$), we shows the ensemble-averaged primary depth as
\begin{equation}
\bar{z}_\mathrm{m} = \frac{1}{n}\sum_{i=1}^{n} z_\mathrm{m}^{(i)},
\qquad
s_{\bar{z}_\mathrm{m}} = \mathrm{SD}\left(z_\mathrm{m}^{(1)},\ldots,z_\mathrm{m}^{(n)}\right),
\end{equation}
where $s_{\bar{z}_\mathrm{m}}$ denotes the standard deviation across the $n$ measurements.

In the following discussion, $z_\mathrm{m}$ (or $\bar{z}_\mathrm{m}$ when averaged) is used as a key descriptor due to being typically most sensitive to the pulse energy-dependent onset of self-focusing-driven modification, whereas the emergence and spread of additional multifocal points are primarily governed by pulse overlap and the associated cumulative effects.

%%       \cite{<label>} ==> [1]




\nolinenumbers

{\newpage}

\end{document}